\documentclass[10pt]{article}

\usepackage[margin=1in]{geometry}
\usepackage{setspace}
\usepackage{amsmath}
\usepackage{amssymb}
\usepackage{bm}

\usepackage{graphicx}
\usepackage{caption}
\usepackage{subcaption}
\usepackage{float}
\usepackage{booktabs}
\usepackage{tabularx}
\usepackage{array}

\usepackage{siunitx}
\usepackage[numbers,sort&compress]{natbib}

\usepackage{xcolor}
\usepackage[
    colorlinks=true,
    linkcolor=blue,
    citecolor=blue,
    urlcolor=blue
]{hyperref}

\title{
Dipole-Field Magnetic Windows for Radio-Frequency Transmission\\
Through Hypersonic Plasma Sheaths: A Reduced-Order Scaling Model
}

\author{Richard Lieu\\
Department of Physics and Astronomy\\
University of Alabama in Huntsville, Huntsville, AL 35899, USA\\
\texttt{lieur@uah.edu}}
\date{}

\begin{document}

\maketitle
\vspace{-0.8em}

\begin{abstract}
Hypersonic vehicles and atmospheric-entry bodies can experience radio-frequency communication blackout when shock-heated gas surrounding the vehicle becomes sufficiently ionized that plasma cutoff and collisional attenuation restrict electromagnetic transmission. Magnetic-window approaches attempt to reduce this loss by exploiting the anisotropic dispersion of a magnetized plasma, in which selected right-hand or whistler-like modes may propagate along preferred directions. This paper develops a reduced-order scaling model for magnetic-window transmission through a finite-thickness hypersonic plasma sheath when the magnetic source is represented as an onboard axial dipole. The model gives a closed-form estimate of the collisionless angular aperture, compares the underlying projected-cyclotron criterion with full cold-plasma dispersion roots, and extends the aperture estimate to a loss-limited cone using a collisional optical-depth approximation. A simplified neutral-density, speed, and ionization closure is used only to generate qualitative parametric maps and sensitivity trends. The results clarify how dipole-field decay, sheath thickness, radio frequency, vehicle scale, electron density, and collisions jointly constrain the candidate transmission window. The contribution is intended as a screening framework for selecting cases for full-wave electromagnetic simulation, nonequilibrium aerothermochemistry, antenna-coupling analysis, and laboratory validation, rather than as a demonstrated engineering solution to plasma blackout.
\end{abstract}

\vspace{0.5em}
\noindent\textbf{Keywords:} hypersonic plasma; communication blackout; magnetized plasma; radio-frequency propagation; dipole field; collisional attenuation

\section{Introduction}

Maintaining radio-frequency links to vehicles in hypersonic flight or atmospheric entry is a persistent aerospace challenge. During high-speed flight through an atmosphere, the gas surrounding a vehicle is compressed and heated in the shock layer, producing dissociation, ionization, and a partially ionized plasma sheath. If the electron number density is sufficiently high, radio-frequency signals used for telemetry, tracking, sensing, or command links may be reflected, absorbed, or converted into evanescent fields before reaching the exterior flow. In the simplest unmagnetized cold-plasma model,
\begin{equation}
    k^2 c^2 = \omega^2 - \omega_p^2,
    \qquad
    \omega_p^2 = \frac{n_e e^2}{\epsilon_0 m_e},
    \label{eq:unmag_dispersion}
\end{equation}
so that propagation is cut off when $\omega_p > \omega$. This elementary criterion does not capture the spatially varying, collisional, chemically reacting, and nonequilibrium nature of a realistic hypersonic plasma sheath, but it identifies electron density as a central parameter in blackout analysis \cite{Rybak1971, Kim2009Review, StanfordReentry}.

A range of blackout-mitigation concepts has been investigated, including operation at higher radio frequency, aerodynamic shaping, material or particulate injection, crossed-field plasma manipulation, and externally imposed magnetic fields. Magnetic-field methods are attractive because a magnetized plasma is anisotropic: the Lorentz force changes the electron response transverse to the magnetic field and permits propagation behavior that is not available in an unmagnetized plasma. Prior analyses of magnetized reentry plasma have shown that magnetic windows can occur through right-hand or whistler-like propagation and that transmission, reflection, and absorption depend strongly on magnetic-field strength, plasma density, and collision frequency \cite{Manning2009}. More detailed numerical studies have examined radio blackout and magnetic-window mitigation in hypersonic flowfields \cite{Kundrapu2015}, while recent work has considered both dipole-field and pulsed-field approaches for manipulating transmission through plasma sheaths \cite{Bai2022EHF, Yuan2021Pulsed, Peng2025Pulsed}.

The present paper does not claim that the magnetic-window mechanism itself is new. Instead, it addresses a reduced design and scaling question that arises when the magnetic source is carried by the flight body and approximated as an axial dipole. This distinction is important for compact or vehicle-integrated concepts because a dipole field decays rapidly with distance. A transmission path that appears open near the surface may close near the outer edge of the sheath. The finite sheath thickness therefore couples directly to vehicle scale, magnetic-field strength, and operating frequency. In the reduced radial-escape model used here, this coupling appears through the dimensionless dipole penalty $(1+t_s/a)^3$, where $t_s$ is the sheath thickness and $a$ is the projectile radius or characteristic nose scale.

The paper develops a compact scaling model for the angular aperture of a dipole-field magnetic window. The collisionless opening angle derived below is given in Eq.~\eqref{eq:theta_open_dimensional}, provided that its inverse-cosine argument is less than unity. This expression is not intended to replace full-wave propagation through a realistic sheath. Its purpose is to expose the leading dependence of the aperture on magnetic-field strength, radio frequency, vehicle scale, and sheath thickness. The same framework is then extended through the collisional optical-depth condition in Eq.~\eqref{eq:tau_reduced}, producing the candidate loss-limited aperture in Eq.~\eqref{eq:theta_loss_limited}.

\begin{figure}[t]
    \centering
    \includegraphics[width=0.88\linewidth]{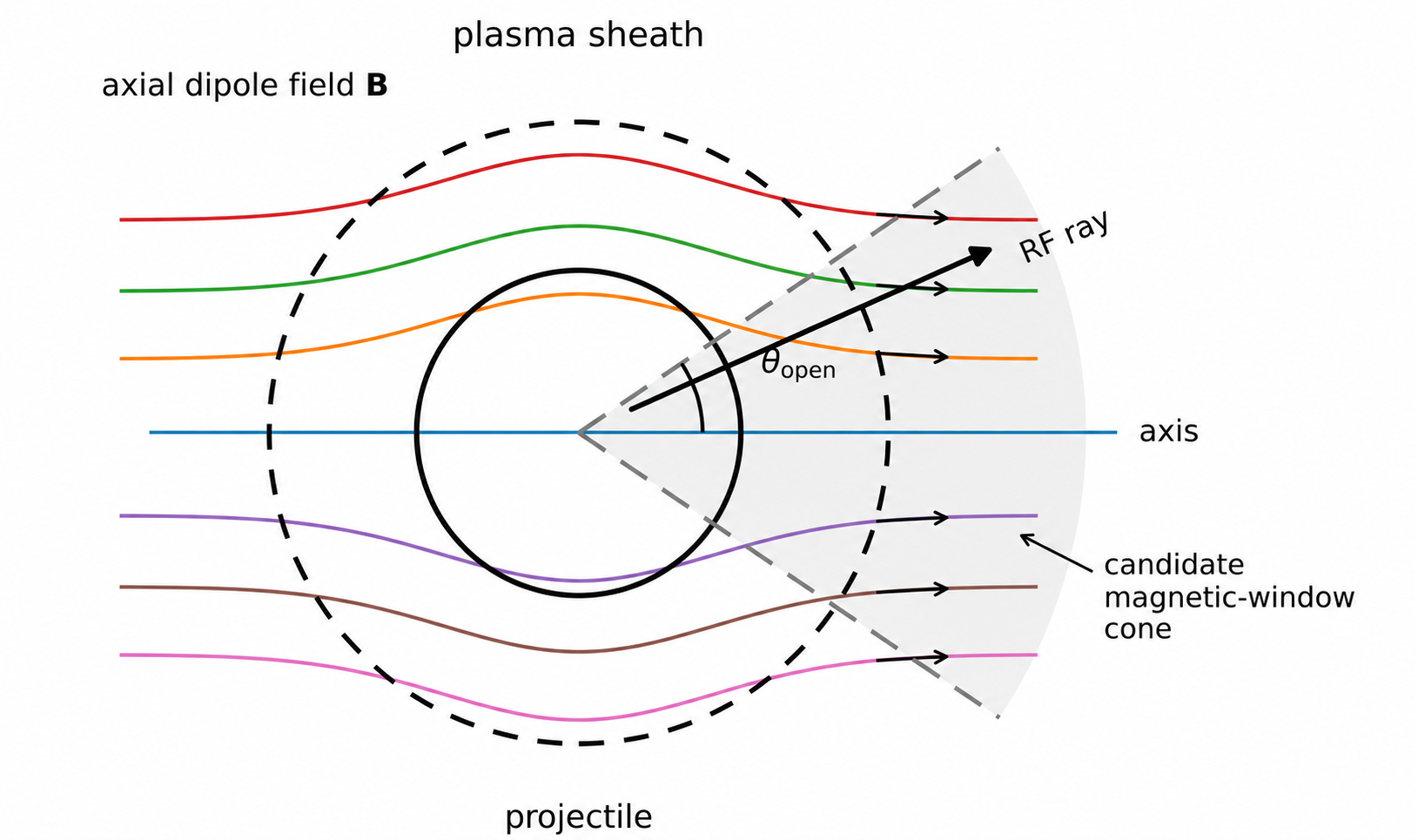}
    \caption{Idealized geometry for a vehicle-borne axial dipole magnetic field interacting with a finite-thickness plasma sheath. The colored curves denote representative axial dipole magnetic-field lines, the dashed curve denotes the outer sheath boundary $r=a+t_s$, and the blue horizontal line denotes the vehicle axis. The lightly shaded wedge denotes the candidate magnetic-window cone. Its two dashed gray boundaries represent the limiting angular directions predicted by the reduced opening-cone model, while the black arrow denotes a representative RF ray inside the cone. The angle $\theta_{\rm open}$ is measured from the vehicle axis to the upper cone boundary, not to the RF ray.}
    \label{fig:dipole-window-schematic}
\end{figure}

Figure~\ref{fig:dipole-window-schematic} illustrates the idealized geometry. The flight body is modeled as carrying an axial dipole field, the plasma sheath is represented as a finite layer surrounding the body, and the shaded wedge denotes the candidate magnetic-window cone predicted by the reduced opening-angle model. In the schematic, the black slanted arrow is a representative RF ray lying inside the cone, whereas the two gray dashed lines are geometric cone boundaries and are not additional propagating rays. The notation used in the derivation is also indicated: the radial coordinate is $r=a+s$, the sheath extends from $s=0$ to $s=t_s$, and the limiting condition is evaluated at the outer sheath boundary.

A second objective is to include collisional loss at a level appropriate for preliminary screening. A magnetic field may create a propagating branch, but it does not remove absorption. In a weakly ionized sheath, electron-neutral collisions can convert wave energy into thermal motion and reduce the transmitted amplitude. The reduced optical-depth expression in Eq.~\eqref{eq:tau_reduced}, with detuning defined by Eq.~\eqref{eq:detuning}, is used only as an attenuation screen; it indicates how plasma density, collision frequency, sheath thickness, and magnetic detuning enter the loss-limited aperture.

The intended contribution is therefore a transparent reduced-order screening model between elementary plasma-cutoff estimates and high-fidelity plasma-flow or electromagnetic simulations. The model is simple enough to be evaluated over wide parameter ranges, yet explicit enough to show when a dipole-field magnetic window is closed by frequency, closed by sheath thickness, opened by magnetic field strength, or rendered ineffective by collisional loss. The resulting scaling relations are intended to guide the selection of cases for full-wave propagation calculations, nonequilibrium aerothermochemical modeling, antenna-coupling analysis, and laboratory validation.

The manuscript is organized to separate reduced theory, derivation, parametric application, and limitations. Sections~3--5 define the modeling assumptions and the unmagnetized and magnetized plasma-wave criteria. Sections~6 and 7 derive the dipole-field aperture and the candidate loss-limited aperture, with detailed algebraic steps retained in the Supplementary Material. Section~8 introduces the reduced sheath closure used only for parametric studies. Section~9 presents the parameter maps, local dispersion-root check, diagnostic trends, and sensitivity tests. Sections~10--13 compare the model with prior work, discuss limitations, and summarize the conclusions.
\section{Position of the Present Study Relative to Prior Magnetic-Window Work}
\label{sec:literature-comparison}

The present study is positioned as a reduced-order scaling analysis rather than as the first demonstration of magnetic-window propagation. Prior work has already established that magnetized plasma can support transmission regimes unavailable in an unmagnetized sheath, especially through right-hand or whistler-like modes. The contribution of the present paper is narrower: it derives an explicit aperture criterion for a projectile-borne axial dipole field, includes the finite-sheath penalty associated with dipole falloff, and introduces a loss-limited cone based on a collisional optical-depth estimate.

Table~\ref{tab:prior-work-comparison} summarizes the relationship between the present model and representative prior studies. The comparison is intended to clarify the novelty claim. The paper does not claim novelty in the basic magnetic-window mechanism. Instead, it contributes a compact analytical connection between surface dipole field, sheath thickness, projectile scale, radio frequency, collisional loss, and angular aperture. This connection is useful because it provides a screening model that can be evaluated before more expensive full-wave, ray-tracing, or nonequilibrium flow simulations are performed.

\begin{table}[t]
    \centering
    \caption{Comparison between the present reduced-order dipole-aperture model and representative prior studies of plasma blackout, magnetic-window mitigation, and dipole-field propagation. The present contribution is not the magnetic-window mechanism itself, but a compact analytical scaling relation for the opening angle and candidate loss-limited cone associated with a projectile-borne axial dipole field.}
    \label{tab:prior-work-comparison}
    \small
    \begin{tabularx}{\linewidth}{>{\raggedright\arraybackslash}p{0.18\linewidth}
                                >{\raggedright\arraybackslash}p{0.24\linewidth}
                                >{\raggedright\arraybackslash}p{0.25\linewidth}
                                >{\raggedright\arraybackslash}X}
        \toprule
        Study & Primary approach & Treatment of magnetic-window physics & Distinction from the present work \\
        \midrule
        Rybak and Churchill \cite{Rybak1971} &
        Review of reentry communication blackout, diagnostic work, and mitigation concepts &
        Establishes the broader plasma-blackout context, including the importance of electron density, plasma frequency, and attenuation mechanisms &
        Provides foundational blackout background rather than a projectile-borne dipole aperture model or loss-limited angular scaling relation \\
        \addlinespace
        Manning \cite{Manning2009} &
        Kinetic-equation analysis of electromagnetic-wave propagation through a magnetized reentry plasma &
        Demonstrates that magnetic windows can arise through whistler-mode propagation and evaluates transmission, reflection, and absorption coefficients for magnetized plasma conditions &
        Treats magnetized-plasma transmission directly, but does not derive the finite-sheath axial-dipole opening-angle formula or the associated projectile-scale dependence emphasized here \\
        \addlinespace
        Kundrapu et al. \cite{Kundrapu2015} &
        Coupled hypersonic-flow and electromagnetic simulation using multispecies flow modeling, multifluid plasma equations, and Maxwell equations &
        Demonstrates blackout over a hypersonic vehicle and examines a magnetic-window mitigation scheme in a first-principles numerical framework &
        Provides higher-fidelity numerical modeling, whereas the present work provides a closed-form screening model for dipole-field aperture and collisional loss \\
        \addlinespace
        Bai et al. \cite{Bai2022EHF} &
        Numerical study of EHF-wave propagation through hypersonic plasma sheaths magnetized by dipole fields generated by coils &
        Directly examines dipole magnetic fields and shows that collisions remain a major dissipation mechanism while dipole fields can differ substantially from uniform-field behavior &
        Closest prior work to the present study; the present paper complements it by deriving a compact analytical opening-cone criterion proportional to $(1+t_s/a)^3$ and a reduced loss-limited cone formula \\
        \bottomrule
    \end{tabularx}
\end{table}

The table also indicates why a reduced model remains useful even though more detailed numerical studies exist. Full simulations can resolve sheath gradients, finite-rate chemistry, wave reflection, and field-plasma interactions, but they may obscure the leading parameter dependencies. The analytical aperture formula derived in this paper isolates the dipole-field penalty through the factor $(1+t_s/a)^3$ in Eq.~\eqref{eq:theta_open_dimensional}, thereby showing how rapidly a projectile-borne field loses effectiveness as the sheath becomes thick relative to the vehicle scale. The loss-limited extension further distinguishes a geometrically open magnetic window from a candidate transmission path that also satisfies an attenuation criterion.

This positioning determines the standard of validation required for the present manuscript. The immediate question is not whether magnetic windows exist, since that has been addressed in prior work. The question is whether the reduced dipole-aperture model provides a transparent and physically consistent screening criterion that agrees in trend with established magnetic-window physics and identifies regimes worth studying with higher-fidelity methods.

\section{Problem Definition and Modeling Assumptions}

This study considers the reduced problem of radio-frequency propagation from a projectile or flight body through a finite-thickness plasma sheath in the presence of a projectile-borne axial dipole magnetic field. The intended setting is a hypersonic or atmospheric-entry flow in which shock heating produces a partially ionized layer surrounding the body. The objective is not to compute the plasma sheath from first principles, but to derive a transparent scaling model for the conditions under which an imposed dipole field can create a directional magnetic window for radio-frequency transmission.

The projectile is idealized as an axisymmetric body with a characteristic radius or nose scale $a$. The plasma sheath is represented as a layer of thickness $t_s$ surrounding the body. The radial coordinate is written as
\begin{equation}
    r = a+s,
    \qquad
    0 \leq s \leq t_s,
\end{equation}
where $s=0$ denotes the nominal vehicle surface and $s=t_s$ denotes the outer edge of the sheath. The polar angle $\theta$ is measured from the projectile axis, which is also taken to be the axis of the magnetic dipole. The principal quantity of interest is the half-angle of a polar region within which radio-frequency energy can escape approximately radially through the sheath.

The imposed magnetic field is assumed to be generated by a compact onboard source and approximated outside the source region as the axial dipole defined later in Eqs.~\eqref{eq:dipole_components} and \eqref{eq:dipole_magnitude}, where $B_p$ is the axial magnetic-field strength at the projectile surface. This vacuum dipole approximation neglects distortion of the magnetic field by induced plasma currents, magnetohydrodynamic modification of the shock layer, and finite-size details of the actual source. These effects are important for a complete vehicle design, but excluding them allows the leading geometric penalty associated with dipole falloff to be isolated.

Radio-frequency rays are assumed to leave the vehicle approximately radially. Under this assumption, the angle $\psi$ between the local propagation direction $\bm{k}$ and the local magnetic field $\bm{B}$ is given by Eq.~\eqref{eq:cospsi_dipole}. This approximation is most appropriate for a first estimate of the polar escape region. A full treatment would require ray tracing or full-wave propagation through a spatially varying anisotropic plasma.

The plasma sheath is characterized locally by an electron number density $n_e$, an effective electron collision frequency $\nu_e$, and an electron plasma frequency $\omega_p$. Unless otherwise stated, these quantities are treated as representative values within the sheath or as functions to be evaluated along a ray path. The plasma is assumed to be weakly magnetized by the vehicle field only in the sense that the imposed field changes the electromagnetic wave response; the background flow, density, and temperature profiles are not solved self-consistently with the magnetic field. The radio wave has angular frequency
\begin{equation}
    \omega = 2\pi f,
\end{equation}
where $f$ is the ordinary frequency.

The modeling hierarchy used in this paper is intentionally reduced. The unmagnetized cutoff condition establishes when a sheath is opaque in the absence of a magnetic field. The magnetized-plasma condition then estimates whether a right-hand or whistler-like branch can propagate along directions sufficiently aligned with the magnetic field. The dipole geometry maps that local propagation condition into an angular opening cone. Finally, a collisional optical-depth estimate reduces the collisionless opening angle to a candidate loss-limited angle. This sequence is designed to produce scaling relations rather than high-fidelity transmission predictions.

Several assumptions should be emphasized because they define the domain of validity of the model. The plasma response is treated using cold-plasma electromagnetic dispersion, with collisional damping added perturbatively. The sheath is treated as sufficiently smooth that local dispersion relations can be used along the path of an escaping wave. The wave amplitude is assumed small enough that the plasma response is linear. The antenna is not modeled explicitly, so coupling efficiency into the relevant magnetized-plasma mode is not included. The chemical state of the gas is represented only through a simplified closure when parametric plots are generated. Nonequilibrium chemistry, radiation, ablation products, turbulence, and detailed shock-layer structure are outside the scope of the present analytical model.

With these assumptions, the central question becomes the following: for a given projectile scale $a$, sheath thickness $t_s$, radio frequency $f$, surface dipole field $B_p$, and plasma state $(n_e,\nu_e)$, does there exist a polar angular interval through which the wave can both propagate and survive collisional attenuation? The following section establishes the baseline unmagnetized cutoff criterion against which the magnetic-window model is compared.

\section{Baseline RF Cutoff in an Unmagnetized Plasma Sheath}

The simplest model of radio-frequency blackout treats the plasma sheath as an unmagnetized cold electron plasma. Ions are assumed to be stationary over the timescale of the radio-frequency oscillation because of their much larger mass, while electrons provide the dominant high-frequency current response. For a harmonic electric field proportional to $\exp(-i\omega t)$, the linearized electron equation of motion in the collisionless limit is
\begin{equation}
    m_e \frac{d\bm{v}}{dt}
    =
    -e\bm{E}.
\end{equation}
For a time-harmonic response, this gives
\begin{equation}
    \bm{v}
    =
    -\frac{i e}{m_e \omega}\bm{E}.
\end{equation}
The corresponding current density is
\begin{equation}
    \bm{J}
    =
    -n_e e\bm{v}
    =
    \frac{i n_e e^2}{m_e \omega}\bm{E}.
\end{equation}
Combining this current response with Maxwell's equations gives the familiar cold-plasma dielectric function
\begin{equation}
    \epsilon(\omega)
    =
    \epsilon_0
    \left(
    1-\frac{\omega_p^2}{\omega^2}
    \right),
\end{equation}
where the electron plasma frequency is
\begin{equation}
    \omega_p
    =
    \left(
    \frac{n_e e^2}{\epsilon_0 m_e}
    \right)^{1/2}.
\end{equation}
The electromagnetic dispersion relation is then
\begin{equation}
    k^2 c^2
    =
    \omega^2-\omega_p^2.
\end{equation}

When $\omega>\omega_p$, the wavenumber $k$ is real and the wave can propagate through the idealized collisionless plasma. When $\omega<\omega_p$, the wavenumber becomes imaginary:
\begin{equation}
    k=i\alpha,
    \qquad
    \alpha
    =
    \frac{1}{c}
    \left(
    \omega_p^2-\omega^2
    \right)^{1/2}.
\end{equation}
The electric field then decays as
\begin{equation}
    E(s)
    =
    E(0)\exp(-\alpha s),
\end{equation}
so that a sheath of thickness $t_s$ transmits an amplitude factor
\begin{equation}
    \frac{E(t_s)}{E(0)}
    =
    \exp(-\alpha t_s).
\end{equation}
Thus, in the absence of a magnetic field, a plasma layer becomes opaque when the local plasma frequency exceeds the radio frequency over a sufficient path length.

The corresponding critical electron density is obtained by setting $\omega_p=\omega$:
\begin{equation}
    n_{e,\mathrm{crit}}
    =
    \frac{\epsilon_0 m_e}{e^2}\omega^2
    =
    \frac{\epsilon_0 m_e}{e^2}(2\pi f)^2.
    \label{eq:necrit_scaling}
\end{equation}
This expression provides a useful first estimate of the electron density required for blackout at a given communication frequency. Frequencies in the gigahertz range require larger electron densities for cutoff than frequencies in the megahertz range, which is why higher-frequency communication links have often been considered as a blackout-mitigation strategy \cite{Kim2009Review, StanfordReentry}.

Collisions modify this simple picture by making the dielectric response complex. If an effective electron collision frequency $\nu_e$ is included in the electron momentum equation,
\begin{equation}
    m_e\frac{d\bm{v}}{dt}
    +
    m_e\nu_e\bm{v}
    =
    -e\bm{E},
\end{equation}
the dielectric function becomes
\begin{equation}
    \epsilon(\omega)
    =
    \epsilon_0
    \left[
    1-
    \frac{\omega_p^2}
    {\omega(\omega+i\nu_e)}
    \right].
\end{equation}
The resulting refractive index is complex, and the wave can be attenuated even when the collisionless cutoff condition is not strictly satisfied. In a hypersonic sheath, this distinction matters because radio-frequency blackout is not only a cutoff phenomenon; it is also an absorption problem. Electron-neutral and electron-ion collisions can dissipate wave energy, while spatial gradients can produce reflection, tunneling, and mode conversion effects not represented in the uniform-plasma expression.

The unmagnetized cutoff criterion nonetheless provides the baseline against which the magnetic-window concept is evaluated. If $n_e<n_{e,\mathrm{crit}}$, the sheath is not overdense at the chosen frequency in the elementary cold-plasma sense, and a magnetic window is not required to avoid cutoff. If $n_e>n_{e,\mathrm{crit}}$, then the unmagnetized sheath is locally overdense, and propagation requires either tunneling through a sufficiently thin layer, operation at a higher frequency, reduction of the electron density, or alteration of the dispersion relation. The remainder of the paper focuses on the last possibility: using an imposed magnetic field to make propagation anisotropic, thereby permitting a directional transmission window through an otherwise opaque sheath \cite{Manning2009, Kundrapu2015, Bai2022EHF}.

\section{Magnetized-Plasma Dispersion and Directional Transmission Criteria}

The unmagnetized cutoff model assumes that the plasma response is isotropic. Once a static magnetic field is imposed, this assumption no longer holds. The Lorentz force constrains electron motion differently along and across the magnetic-field direction, so the dielectric response becomes tensorial. This anisotropy is the basis of the magnetic-window concept. A plasma that is opaque to an ordinary electromagnetic wave in the absence of a magnetic field may support propagation of a particular polarization along directions sufficiently aligned with the imposed field.

Let the background magnetic field be denoted by $\bm{B}_0$, and choose the local $z$-axis to be parallel to $\bm{B}_0$. The electron cyclotron frequency is
\begin{equation}
    \Omega_e
    =
    \frac{eB_0}{m_e},
\end{equation}
where $B_0=|\bm{B}_0|$. In a cold magnetized electron plasma, the dielectric response may be written in the form
\begin{equation}
    \bm{K}
    =
    \begin{pmatrix}
        S & -iD & 0 \\
        iD & S & 0 \\
        0 & 0 & P
    \end{pmatrix},
    \label{eq:cold_tensor}
\end{equation}
where $\bm{K}$ is the relative dielectric tensor. The quantities $S$, $D$, and $P$ are commonly expressed in terms of the right-hand and left-hand circularly polarized responses $R$ and $L$:
\begin{equation}
    S=\frac{R+L}{2},
    \qquad
    D=\frac{R-L}{2},
    \qquad
    P=1-\frac{\omega_p^2}{\omega^2}.
\end{equation}
In the collisionless electron-dominated limit, the circular responses are
\begin{equation}
    R
    =
    1-\frac{\omega_p^2}{\omega(\omega-\Omega_e)},
\end{equation}
and
\begin{equation}
    L
    =
    1-\frac{\omega_p^2}{\omega(\omega+\Omega_e)}.
\end{equation}
The sign convention used here treats $\Omega_e$ as a positive magnitude. With this convention, the right-hand branch contains the resonance at $\omega=\Omega_e$ and is the branch relevant to the whistler-like magnetic-window behavior considered in this paper.

For a plane wave with refractive index $n=ck/\omega$, Maxwell's equations may be written as
\begin{equation}
    \left[
    n^2\bm{I}
    -
    \bm{n}\bm{n}
    -
    \bm{K}
    \right]\cdot \bm{E}
    =
    0,
\end{equation}
where $\bm{n}$ is the refractive-index vector. Nontrivial wave solutions require the determinant of this system to vanish. The roots discussed below are therefore the two allowed values of $n^2$, i.e. the squared refractive indices of the local electromagnetic normal modes. They are not independent scalar-permittivity roots. If the wave vector makes an angle $\psi$ with the magnetic field, the cold-plasma dispersion relation may be written as a quadratic equation in $n^2$:
\begin{equation}
    A n^4 - B n^2 + C = 0,
    \label{eq:cold_quadratic}
\end{equation}
where
\begin{equation}
    A
    =
    S\sin^2\psi+P\cos^2\psi,
    \label{eq:cold_A}
\end{equation}
\begin{equation}
    B
    =
    RL\sin^2\psi+PS(1+\cos^2\psi),
    \label{eq:cold_B}
\end{equation}
and
\begin{equation}
    C
    =
    PRL.
    \label{eq:cold_C}
\end{equation}
This expression is the local cold-plasma form underlying the Appleton--Hartree dispersion relation. It shows explicitly that propagation is no longer determined only by $\omega_p$ and $\omega$, but also by the angle between the wave vector and the magnetic field.

The most transparent limiting case occurs for propagation parallel to the magnetic field, for which $\psi=0$. The two normal modes are then circularly polarized and have refractive indices
\begin{equation}
    n_R^2
    =
    1-\frac{\omega_p^2}{\omega(\omega-\Omega_e)}
\end{equation}
and
\begin{equation}
    n_L^2
    =
    1-\frac{\omega_p^2}{\omega(\omega+\Omega_e)}.
\end{equation}
The left-hand mode remains subject to cutoff-like behavior in the overdense regime. The right-hand mode behaves differently when the cyclotron frequency exceeds the wave frequency. If $\Omega_e>\omega$, then $\omega-\Omega_e<0$ and the right-hand refractive index becomes
\begin{equation}
    n_R^2
    =
    1+
    \frac{\omega_p^2}{\omega(\Omega_e-\omega)}.
\end{equation}
This quantity is positive even when $\omega_p>\omega$. Thus, for propagation parallel to the magnetic field, a magnetized plasma can support a right-hand mode under conditions that would be cutoff in the corresponding unmagnetized plasma. This is the simplest mathematical statement of the magnetic-window mechanism \cite{Manning2009, Kundrapu2015, Bai2022EHF}.

For oblique propagation, the same physical idea survives in approximate form. In the overdense, low-frequency, electron-dominated regime associated with the whistler-like branch, the effective propagation condition may be expressed as
\begin{equation}
    \Omega_e \cos\psi > \omega.
    \label{eq:projected_cyclotron_condition}
\end{equation}
This relation should be understood as a reduced directional criterion rather than a replacement for the full tensor dispersion relation. It captures the fact that the component of electron gyromotion relevant to wave propagation along the ray direction must be sufficiently large to keep the branch propagating. Equivalently, the local magnetic field must be strong enough, and the ray sufficiently aligned with the field, for the right-hand branch to avoid cutoff.

If collisions are neglected, the local angular condition may be written as
\begin{equation}
    \cos\psi
    >
    \frac{\omega}{\Omega_e}.
\end{equation}
A magnetic window therefore exists only where $\Omega_e>\omega$. When this inequality is barely satisfied, the angular interval around the magnetic-field direction is narrow. When $\Omega_e$ substantially exceeds $\omega$, the angular interval widens. In a uniform magnetic field, this would correspond to a cone of allowed propagation directions around $\bm{B}_0$ with approximate half-angle
\begin{equation}
    \psi_{\rm open}
    =
    \cos^{-1}
    \left(
    \frac{\omega}{\Omega_e}
    \right).
\end{equation}
For a projectile-borne dipole field, however, both the magnitude and direction of $\bm{B}_0$ vary through the sheath. The opening angle measured from the projectile axis is therefore not simply $\psi_{\rm open}$; it must be obtained by combining the local magnetized-plasma criterion with the dipole-field geometry.

The magnetic-window condition should not be interpreted as eliminating dispersion. The right-hand branch remains dispersive, and in the whistler regime the phase and group velocities depend strongly on frequency, magnetic-field strength, and plasma density. The magnetic field changes the dispersion relation so that a propagating branch exists in selected directions; it does not make the plasma equivalent to vacuum. The condition should also not be interpreted as eliminating absorption. In a collisional sheath, the refractive index becomes complex, and a wave that is allowed by the real part of the dispersion relation may still be attenuated strongly before reaching free space.

To include collisions at a reduced level, the parallel right-hand response may be written approximately as
\begin{equation}
    n_R^2
    \simeq
    1-
    \frac{\omega_p^2}
    {\omega(\omega+i\nu_e-\Omega_e)}.
\end{equation}
Introducing the magnetic detuning
\begin{equation}
    \Delta
    =
    \Omega_e\cos\psi-\omega,
    \label{eq:detuning}
\end{equation}
the corresponding oblique reduced form used later in this paper is
\begin{equation}
    n^2
    \simeq
    1+
    \frac{\omega_p^2}{\omega(\Delta-i\nu_e)}.
\end{equation}
This expression separates the geometric opening condition from the loss mechanism. A positive detuning favors propagation, whereas a finite collision frequency introduces an imaginary part into the refractive index. The ratio of $\nu_e$ to $\Delta$ therefore helps determine whether the magnetic window is merely open in the collisionless sense or a candidate for communication-system analysis.

The following section applies this local propagation criterion to the specific magnetic-field geometry of an axial dipole carried by the projectile. The result is a closed-form expression for the collisionless opening angle and a natural basis for the loss-limited cone derived later.

\section{Axial Dipole Geometry and Collisionless Opening-Cone Criterion}
\label{sec:dipole_opening}

The projectile-borne magnetic source is approximated as an axial dipole whose moment is aligned with the vehicle symmetry axis. Let $a$ denote the projectile radius or characteristic nose scale, $B_p$ the magnetic-field strength at the surface on the axis, $r$ the distance from the dipole center, and $\theta$ the polar angle from the axis. Outside the source region, the dipole components are written as
\begin{equation}
    B_r(r,\theta)=B_p\left(\frac{a}{r}\right)^3\cos\theta,
    \qquad
    B_\theta(r,\theta)=\frac{B_p}{2}\left(\frac{a}{r}\right)^3\sin\theta .
    \label{eq:dipole_components}
\end{equation}
The corresponding field magnitude is
\begin{equation}
    B(r,\theta)
    =
    \frac{B_p}{2}\left(\frac{a}{r}\right)^3
    \left(1+3\cos^2\theta\right)^{1/2} .
    \label{eq:dipole_magnitude}
\end{equation}
This vacuum dipole approximation is used only to isolate the geometric $r^{-3}$ field decay. It does not include plasma-current distortion of the field or magnetohydrodynamic feedback on the shock layer.

The RF path is assumed to leave the vehicle approximately radially. The angle $\psi$ between the ray direction and the local magnetic field therefore satisfies
\begin{equation}
    \cos\psi=\frac{\bm{B}\cdot\hat{\bm r}}{B}=\frac{B_r}{B}
    =
    \frac{2\cos\theta}{(1+3\cos^2\theta)^{1/2}} .
    \label{eq:cospsi_dipole}
\end{equation}
The local projected-cyclotron condition from Section~5 is
\begin{equation}
    \Omega_e\cos\psi>\omega,
    \qquad
    \Omega_e=\frac{eB}{m_e} .
\end{equation}
The geometry and plasma condition reduce to the following chain:
\begin{equation}
    \omega
    <
    \Omega_e\cos\psi
    =
    \frac{eB}{m_e}\frac{B_r}{B}
    =
    \frac{eB_r}{m_e}
    =
    \frac{eB_p}{m_e}
    \left(\frac{a}{r}\right)^3\cos\theta .
    \label{eq:projected_chain}
\end{equation}
Thus, for the assumed radial escape path, the relevant magnetic-field component is the radial component of the axial dipole at the point where the ray crosses the sheath. Rearranging Eq.~\eqref{eq:projected_chain} gives the local angular constraint
\begin{equation}
    \cos\theta>
    \frac{m_e\omega}{eB_p}\left(\frac{r}{a}\right)^3 .
    \label{eq:local_theta_constraint}
\end{equation}
The most restrictive point along a radial path is the outer edge of the sheath, $r=a+t_s$, because the dipole field decreases monotonically as $r^{-3}$. At the boundary of the candidate cone, the inequality becomes an equality,
\begin{equation}
    \cos\theta_{\rm open}
    =
    \frac{m_e\omega}{eB_p}
    \left(1+\frac{t_s}{a}\right)^3 .
\end{equation}
The collisionless opening half-angle is therefore
\begin{equation}
    \theta_{\rm open}
    =
    \cos^{-1}\left[
    \frac{m_e\omega}{eB_p}
    \left(1+\frac{t_s}{a}\right)^3
    \right]
    =
    \cos^{-1}\left[
    \frac{m_e2\pi f}{eB_p}
    \left(1+\frac{t_s}{a}\right)^3
    \right] .
    \label{eq:theta_open_dimensional}
\end{equation}
This is the central collisionless screening relation. It is defined only when the inverse-cosine argument is less than unity. The corresponding minimum axial surface field is
\begin{equation}
    B_{p,\min}
    =
    \frac{m_e2\pi f}{e}
    \left(1+\frac{t_s}{a}\right)^3 .
    \label{eq:Bmin}
\end{equation}
The factor $(1+t_s/a)^3$ is the finite-sheath penalty associated with carrying the magnetic source on the vehicle. A uniform-field estimate would require a field of order $m_e\omega/e$, but a vehicle-borne dipole must maintain that effective field at the outer sheath boundary rather than only at the surface.

It is useful to introduce the dimensionless surface magnetization and sheath-thickness parameters
\begin{equation}
    Y_p=\frac{eB_p}{m_e\omega},
    \qquad
    \Lambda=\frac{t_s}{a} .
    \label{eq:Yp_Lambda}
\end{equation}
Equation~\eqref{eq:theta_open_dimensional} becomes
\begin{equation}
    \theta_{\rm open}
    =
    \cos^{-1}\left[\frac{(1+\Lambda)^3}{Y_p}\right],
    \label{eq:theta_open_dimensionless}
\end{equation}
with existence condition
\begin{equation}
    Y_p>(1+\Lambda)^3 .
    \label{eq:window_existence}
\end{equation}
Increasing $B_p$ or decreasing $f$ increases $Y_p$ and opens the aperture. Increasing the relative sheath thickness $\Lambda$ closes the aperture rapidly because of the cubic dipole decay.

Near threshold, let $Y_p=(1+\Lambda)^3(1+\epsilon)$ with $0<\epsilon\ll1$. Then $(1+\Lambda)^3/Y_p\simeq1-\epsilon$ and, using $\cos\theta\simeq1-\theta^2/2$,
\begin{equation}
    \theta_{\rm open}\simeq(2\epsilon)^{1/2} .
    \label{eq:near_threshold_angle}
\end{equation}
Thus a field only slightly above the opening threshold produces a narrow aperture. The collisionless opening angle is therefore a necessary geometric condition, not a sufficient condition for a communication link. Collisional damping, reflection from gradients, ray bending, polarization mismatch, antenna coupling, and link-budget constraints can all reduce or eliminate the candidate transmission region.

\section{Collisional Damping and Candidate Loss-Limited Aperture}
\label{sec:loss_limited_aperture}

The collisionless aperture identifies directions in which the reduced magnetized-plasma branch can propagate in principle. It does not determine whether a signal remains large enough after crossing a collisional sheath. For a right-hand or whistler-like branch sufficiently close to the magnetic-field direction, the collisionless denominator measures the separation from the projected cyclotron boundary. Collisions are introduced in the usual cold-plasma manner by adding an effective damping rate to the electron response. With the magnetic detuning defined in Eq.~\eqref{eq:detuning}, the reduced collisional response is written as
\begin{equation}
    n^2\simeq1+\frac{\omega_p^2}{\omega(\Delta-i\nu_e)},
    \label{eq:n2_collisional}
\end{equation}
where $\nu_e$ is an effective electron collision frequency. This is the projected-direction analogue of the standard collisional right-hand cold-plasma response \cite{Stix1992,Swanson2003}. Positive detuning corresponds to the collisionless propagating side of the projected-cyclotron condition, while finite $\nu_e$ introduces absorption.

Writing $n=n_r+in_i$ and assuming weak damping, $n_i\ll n_r$, gives
\begin{equation}
    n_i\simeq
    \frac{1}{2n_r}
    \frac{\omega_p^2}{\omega}
    \frac{\nu_e}{\Delta^2+\nu_e^2} .
    \label{eq:ni_approx}
\end{equation}
The transmitted amplitude is represented by $\exp(-\tau)$, where
\begin{equation}
    \tau=\frac{\omega}{c}\int_0^{t_s} n_i(s)\,ds .
    \label{eq:tau_integral}
\end{equation}
The reduced propagation condition enters this path integral through $\Delta(s)=\Omega_e(s)\cos\psi(s)-\omega$ inside $n_i(s)$: locations close to $\Delta=0$ are more strongly damped, whereas larger positive detuning lowers the weak-damping loss term. Using representative sheath properties gives the optical-depth screening relation
\begin{equation}
    \tau\simeq
    \frac{t_s}{2cn_r}
    \omega_p^2
    \frac{\nu_e}{\Delta^2+\nu_e^2} .
    \label{eq:tau_reduced}
\end{equation}
This expression is not a substitute for profile-resolved full-wave propagation. It is used only to track how electron density, collision frequency, detuning, and path length affect the candidate aperture.

For an assigned tolerance $\tau_*$, Eq.~\eqref{eq:tau_reduced} gives the minimum detuning required by the loss screen,
\begin{equation}
    \Delta_{\min}
    =
    \left[
    \frac{t_s\omega_p^2\nu_e}{2cn_r\tau_*}
    -\nu_e^2
    \right]^{1/2},
    \label{eq:delta_min}
\end{equation}
with $\Delta_{\min}=0$ if the term inside the square root is negative. At the outer edge of an axial dipole sheath,
\begin{equation}
    \Delta_s(\theta)
    =
    \frac{eB_p}{m_e}
    \left(1+\frac{t_s}{a}\right)^{-3}\cos\theta-
    \omega .
    \label{eq:outer_detuning}
\end{equation}
Setting $\Delta_s(\theta_*)=\Delta_{\min}$ gives the candidate loss-limited half-angle
\begin{equation}
    \theta_*
    =
    \cos^{-1}\left[
    \frac{m_e}{eB_p}
    (\omega+\Delta_{\min})
    \left(1+\frac{t_s}{a}\right)^3
    \right].
    \label{eq:theta_loss_limited}
\end{equation}
When $\Delta_{\min}=0$, Eq.~\eqref{eq:theta_loss_limited} reduces to the collisionless aperture. When collisional damping is significant, $\Delta_{\min}>0$ and the candidate loss-limited aperture is smaller.

The scaling in Eq.~\eqref{eq:tau_reduced} explains the main trends used later. Larger $n_e$ increases $\omega_p^2$ and strengthens the plasma response. Larger $\nu_e$ increases the dissipative channel. Larger $t_s$ both lengthens the absorbing path and weakens the dipole field at the sheath boundary. Larger $B_p$ can widen the collisionless aperture and increase detuning away from the lossy boundary. Since antenna coupling, reflection, ray bending, polarization matching, and a full link budget are not included, $\theta_*$ is interpreted only as a candidate loss-limited transmission aperture.

\section{Plasma-Sheath Scaling Model for Parametric Studies}

The analytical opening-cone and loss-limited transmission formulas derived above require estimates of the electron density, neutral density, collision frequency, and sheath thickness. In a complete treatment, these quantities would be obtained from nonequilibrium hypersonic-flow simulation coupled to finite-rate chemistry, radiation, ablation products if present, and electromagnetic-field effects. The purpose of the present section is more limited. A reduced plasma-sheath closure is introduced only to support parametric scaling studies and to identify qualitative regimes in which a dipole-field magnetic window may be possible, collisionally suppressed, or unnecessary because the unmagnetized sheath is already transmissive.

The upstream neutral number density is denoted by $n_\infty$, and the projectile speed by $V$. The gas immediately behind the shock is represented by an effective temperature
\begin{equation}
    T_s
    =
    T_\infty
    +
    \eta\frac{V^2}{2c_p},
    \label{eq:Ts_model}
\end{equation}
where $T_\infty$ is the upstream temperature, $c_p$ is an effective constant-pressure specific heat, $V$ is the projectile speed, and $\eta$ is a dimensionless factor that accounts, in a lumped way, for the fact that not all directed kinetic energy appears as translational temperature available for ionization. The parameter $\eta$ therefore represents real-gas losses, dissociation, radiation, nonequilibrium effects, and other physics omitted from the reduced model. It should not be regarded as a universal constant.

The post-shock total number density is approximated as
\begin{equation}
    n_s
    =
    C_s n_\infty,
    \label{eq:postshock_density}
\end{equation}
where $C_s$ is an effective compression factor. For an ideal strong shock in a monatomic gas this factor would be finite and of order unity, but in the present model $C_s$ is retained as an adjustable parameter because the actual compression in a hypersonic plasma sheath depends on thermochemistry, geometry, and altitude.

The ionization fraction is denoted by $\chi$, so that
\begin{equation}
    n_e
    =
    \chi n_s
    =
    \chi C_s n_\infty,
    \label{eq:ne_chi}
\end{equation}
and the remaining neutral density is
\begin{equation}
    n_n
    =
    (1-\chi)n_s
    =
    (1-\chi)C_s n_\infty.
    \label{eq:nn_chi}
\end{equation}
For the scaling calculations, $\chi$ is estimated using a simplified Saha-type equilibrium closure. Let $E_i$ be an effective ionization energy representative of the gas mixture. The Saha factor is written as
\begin{equation}
    K(T_s)
    =
    g
    \left(
    \frac{2\pi m_e k_B T_s}{h^2}
    \right)^{3/2}
    \exp
    \left(
    -\frac{E_i}{k_B T_s}
    \right),
    \label{eq:saha_factor}
\end{equation}
where $g$ is an effective statistical-weight factor, $k_B$ is Boltzmann's constant, $h$ is Planck's constant, and $E_i$ is the effective ionization energy. The ionization fraction satisfies
\begin{equation}
    \frac{\chi^2}{1-\chi}n_s
    =
    K(T_s).
    \label{eq:saha_chi}
\end{equation}
In the weak-ionization limit, where $\chi\ll 1$, this reduces to
\begin{equation}
    \chi
    \simeq
    \left[
    \frac{K(T_s)}{C_s n_\infty}
    \right]^{1/2}.
    \label{eq:weak_chi}
\end{equation}
The electron density is then
\begin{equation}
    n_e
    \simeq
    \left[
    K(T_s) C_s n_\infty
    \right]^{1/2}.
    \label{eq:ne_weak}
\end{equation}
This expression shows the competing nature of density and temperature in the reduced closure. At fixed speed, the electron density increases with the square root of the upstream density in the weak-ionization limit. At fixed density, the electron density is extremely sensitive to speed through the exponential dependence of $K(T_s)$ on $T_s$.

The plasma frequency used in the cutoff and damping models is then
\begin{equation}
    \omega_p^2
    =
    \frac{n_e e^2}{\epsilon_0 m_e}.
    \label{eq:omega_p_scaling}
\end{equation}
The baseline unmagnetized blackout condition is $\omega_p>\omega$, or equivalently the density threshold in Eq.~\eqref{eq:necrit_scaling}.
If this condition is not satisfied, the reduced model classifies the sheath as not locally overdense at the radio frequency under consideration. A magnetic window may still influence attenuation or propagation details, but it is not required to overcome elementary plasma cutoff.

The effective collision frequency is estimated using an electron-neutral momentum-transfer model,
\begin{equation}
    \nu_e
    =
    n_n \sigma_{en} v_e,
    \label{eq:nu_model}
\end{equation}
where $\sigma_{en}$ is an effective electron-neutral momentum-transfer cross section and $v_e$ is a characteristic electron thermal speed. In the absence of a separate electron-energy equation, the electron temperature is approximated by the same effective sheath temperature used in the ionization closure, so that
\begin{equation}
    v_e
    =
    \left(
    \frac{8k_B T_s}{\pi m_e}
    \right)^{1/2}.
    \label{eq:ve_model}
\end{equation}
Substitution gives
\begin{equation}
    \nu_e
    =
    (1-\chi)C_s n_\infty \sigma_{en}
    \left(
    \frac{8k_B T_s}{\pi m_e}
    \right)^{1/2}.
    \label{eq:nu_scaling}
\end{equation}
This form captures the expected increase of collisional damping with neutral density and thermal speed. It also captures the reduction of electron-neutral collisions as the gas approaches full ionization, although electron-ion collisions are not explicitly included in the present reduced model.

The loss-limited cone formula requires an estimate of the real part of the refractive index. For parametric screening, the model uses a representative value $n_r$ obtained from the magnitude of the magnetized branch or, in simplified plots, from a bounded estimate based on the dimensionless plasma parameter
\begin{equation}
    X
    =
    \frac{\omega_p^2}{\omega^2}.
    \label{eq:X_param}
\end{equation}
The precise choice of $n_r$ affects the numerical value of the optical depth but not the main qualitative trends. In a high-fidelity calculation, $n_r$ would be obtained from the full complex Appleton--Hartree relation evaluated along the ray path.

The final reduced transmission model combines the plasma closure with the dipole-field damping criterion. The collisionless opening angle, the additional detuning requirement, and the loss-limited angle are given by Eqs.~\eqref{eq:theta_open_dimensional}, \eqref{eq:delta_min}, and \eqref{eq:theta_loss_limited}, respectively. The parameter $\tau_*$ is the maximum acceptable optical depth. In this model, the magnetic window is considered collisionlessly open when Eq.~\eqref{eq:theta_open_dimensional} is real, and it is classified as a candidate loss-limited aperture at the specified optical-depth tolerance when Eq.~\eqref{eq:theta_loss_limited} is real.

The reduced closure should be interpreted cautiously. It does not resolve nonequilibrium vibrational excitation, finite-rate dissociation, multiple ion species, ablation contaminants, sheath curvature, plasma turbulence, or electromagnetic feedback on the flow. Its role is to translate the analytical magnetic-window formulas into interpretable trends with respect to $n_\infty$, $V$, $a$, $t_s$, $B_p$, and $f$. The resulting plots should therefore be read as scaling maps rather than quantitative flight predictions.

\section{Results}

The results presented here are intended to illustrate the behavior of the analytical model rather than to predict a specific vehicle or trajectory. The baseline parameter choices are selected to place the model in a representative hypersonic-plasma communication regime, with a finite sheath surrounding a projectile-scale body and a projectile-borne dipole field strong enough to produce a magnetic-window cone over part of the explored range. Unless otherwise stated, the projectile scale is $a=0.10~\mathrm{m}$, the sheath thickness is $t_s=0.03~\mathrm{m}$, and the nominal radio frequency is $f=2~\mathrm{GHz}$. The field strength $B_p$, ambient neutral density $n_\infty$, and projectile speed $V$ are then varied to expose the dominant dependencies.

The plots in this section are generated directly from the formulas derived above. The collisionless results depend only on the dimensionless magnetization parameter $Y_p=eB_p/(m_e\omega)$ and the relative sheath thickness $\Lambda=t_s/a$. The lossy results additionally depend on the reduced plasma-sheath closure for $n_e$ and $\nu_e$, and therefore on the assumed ionization model, collision cross section, and optical-depth tolerance. The collisionless plots should therefore be regarded as more robust than the loss-limited plots, while the latter are used only to identify candidate loss-limited apertures.

Table~\ref{tab:baseline-parameters} lists the baseline numerical parameters used in the parametric calculations. Table~\ref{tab:figure-parameters} lists the figure-specific sweeps and the parameters held fixed. These tables are intended to make the plotted results reproducible from the equations in Sections~6--8 and the implementation notes in the Supplementary Material.

\begin{table}[t]
    \centering
    \caption{Baseline parameters used in the parametric calculations unless otherwise stated. The values define a screening case and are not intended as a validated flight condition.}
    \label{tab:baseline-parameters}
    \small
    \begin{tabularx}{\linewidth}{>{\raggedright\arraybackslash}p{0.29\linewidth} >{\raggedright\arraybackslash}p{0.20\linewidth} X}
        \toprule
        Quantity & Baseline value & Role in the reduced model \\
        \midrule
        Projectile scale $a$ & $0.10~\mathrm{m}$ & Length scale for the dipole falloff and $\Lambda=t_s/a$ \\
        Sheath thickness $t_s$ & $0.03~\mathrm{m}$ & Path length and outer-boundary field penalty \\
        Radio frequency $f$ & $2.0~\mathrm{GHz}$ & Wave frequency unless varied \\
        Axial surface field $B_p$ & $0.50~\mathrm{T}$ & Dipole field strength unless varied \\
        Projectile speed $V$ & $7.5~\mathrm{km\,s^{-1}}$ & Effective post-shock temperature unless varied \\
        Upstream density $n_\infty$ & $10^{21}~\mathrm{m^{-3}}$ & Neutral-density closure unless varied \\
        Upstream temperature $T_\infty$ & $220~\mathrm{K}$ & Temperature offset in Eq.~\eqref{eq:Ts_model} \\
        Specific heat $c_p$ & $1004~\mathrm{J\,kg^{-1}\,K^{-1}}$ & Effective shock-temperature model \\
        Heating factor $\eta$ & $0.18$ & Lumped real-gas and nonequilibrium correction \\
        Compression factor $C_s$ & $4.0$ & Post-shock number-density multiplier \\
        Effective ionization energy $E_i$ & $14.5~\mathrm{eV}$ & Saha-type closure parameter \\
        Saha degeneracy factor $g$ & $2.0$ & Saha-type closure parameter \\
        Momentum-transfer cross section $\sigma_{en}$ & $10^{-19}~\mathrm{m^2}$ & Electron-neutral collision estimate \\
        Optical-depth tolerance $\tau_*$ & $1.0$ & Candidate loss-limited aperture criterion \\
        Refractive-index estimate $n_r$ & $[\max(X,1)]^{1/2}$ & Representative value for the optical-depth model \\
        \bottomrule
    \end{tabularx}
\end{table}

\begin{table}[t]
    \centering
    \caption{Figure-specific parameter sweeps. Parameters not listed in the varied column are held at the baseline values in Table~\ref{tab:baseline-parameters}.}
    \label{tab:figure-parameters}
    \scriptsize
    \begin{tabularx}{\linewidth}{>{\raggedright\arraybackslash}p{0.13\linewidth} >{\raggedright\arraybackslash}p{0.33\linewidth} X}
        \toprule
        Figure & Parameters varied & Fixed or special assumptions \\
        \midrule
        Fig.~\ref{fig:dipole-window-schematic} & Schematic only & Not a quantitative calculation \\
        Fig.~\ref{fig:angle-vs-Bp} & $B_p=0.03$--$1.5~\mathrm{T}$; curves for $f=0.5$, $1$, $2$, and $5~\mathrm{GHz}$ & Collisionless aperture; $a=0.10~\mathrm{m}$ and $t_s=0.03~\mathrm{m}$ \\
        Fig.~\ref{fig:appleton-validation} & $Y=1.02$--$6$; curves for $X=5$, $20$, and $100$ & Uniform, collisionless cold-plasma roots; $\psi=0$--$89.9^\circ$ \\
        Fig.~\ref{fig:lossy-angle-vs-density} & $n_\infty=10^{18}$--$10^{23}~\mathrm{m^{-3}}$; curves for $B_p=0.2$, $0.4$, $0.8$, and $1.2~\mathrm{T}$ & $V=7.5~\mathrm{km\,s^{-1}}$, $f=2~\mathrm{GHz}$ \\
        Fig.~\ref{fig:lossy-angle-vs-speed} & $V=3$--$12~\mathrm{km\,s^{-1}}$; curves for $n_\infty=10^{19}$, $10^{20}$, $10^{21}$, and $10^{22}~\mathrm{m^{-3}}$ & $B_p=0.50~\mathrm{T}$, $f=2~\mathrm{GHz}$ \\
        Fig.~\ref{fig:speed-closure-diagnostics} & $V=3$--$12~\mathrm{km\,s^{-1}}$ & Closure diagnostics shown for $n_\infty=10^{21}~\mathrm{m^{-3}}$ \\
        Fig.~\ref{fig:angle-vs-sheath-ratio} & $\Lambda=t_s/a=0.01$--$1.5$; curves for $B_p=0.2$, $0.5$, and $1.0~\mathrm{T}$ & Collisionless aperture; $f=2~\mathrm{GHz}$ \\
        Fig.~\ref{fig:regime-map-Bp-density} & $B_p=0.05$--$1.5~\mathrm{T}$ and $n_\infty=10^{18}$--$10^{23}~\mathrm{m^{-3}}$ & Regime classification at $V=7.5~\mathrm{km\,s^{-1}}$ and $f=2~\mathrm{GHz}$ \\
        Fig.~\ref{fig:sensitivity-closure-parameters} & Multipliers $0.5$--$2.0$ applied to $\eta$, $\sigma_{en}$, and $t_s/a$ & Local sensitivity test about the baseline case \\
        \bottomrule
    \end{tabularx}
\end{table}

\subsection{Collisionless Opening-Cone Scaling}

The collisionless opening angle is governed by Eq.~\eqref{eq:theta_open_dimensionless}, with $Y_p$ and $\Lambda$ defined in Eq.~\eqref{eq:Yp_Lambda}. This expression contains the central geometric and electromagnetic scaling of the dipole model. For fixed projectile and sheath dimensions, increasing the surface dipole field increases $Y_p$ and widens the polar transmission cap. Increasing the radio frequency decreases $Y_p$ and narrows the cap. Increasing the sheath thickness relative to the projectile scale increases $\Lambda$ and imposes a cubic penalty.

\begin{figure}[t]
    \centering
    \includegraphics[width=0.74\linewidth]{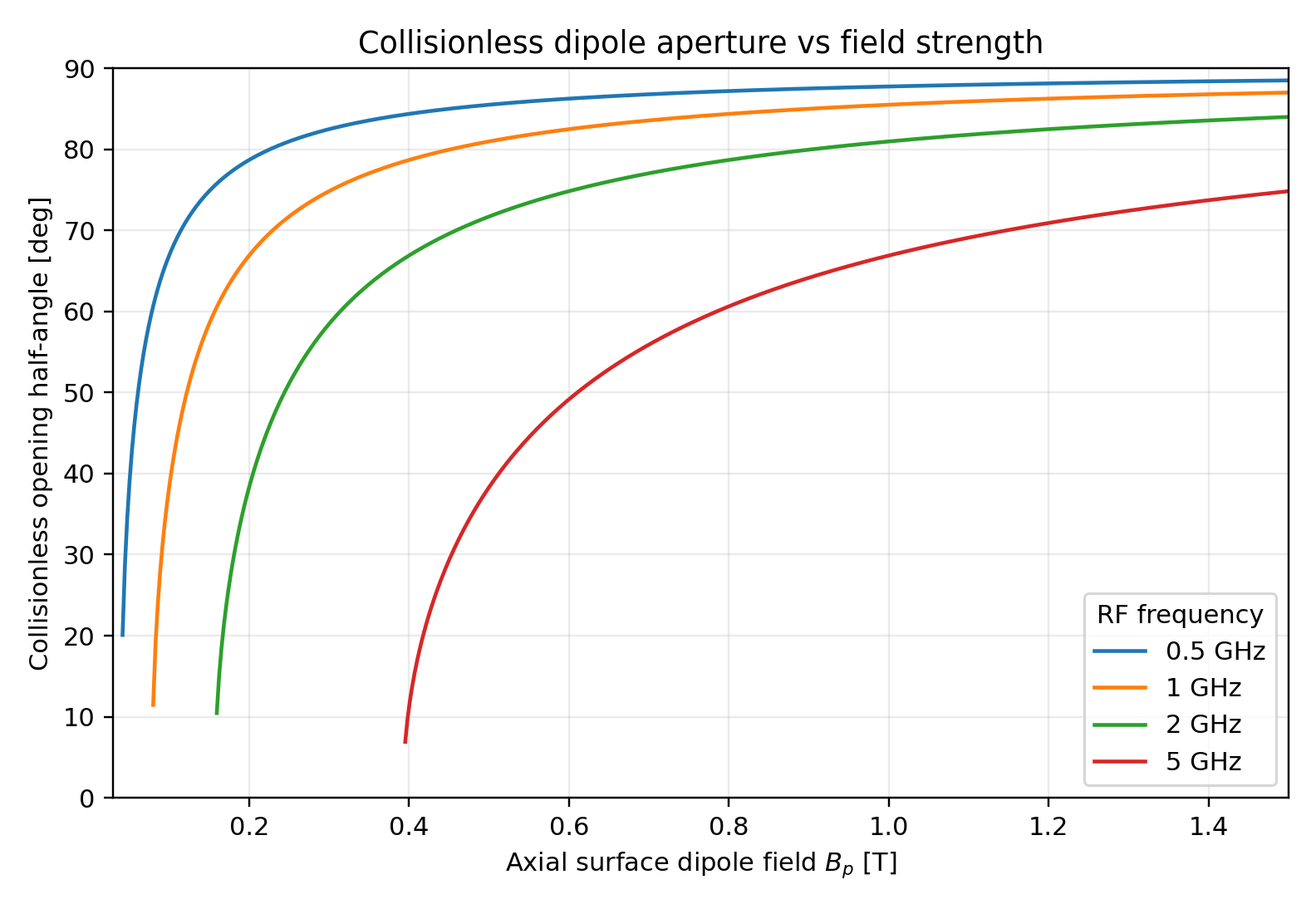}
    \caption{Collisionless magnetic-window opening half-angle as a function of axial surface dipole field $B_p$ for several radio frequencies, using the reduced dipole formula with fixed projectile scale and sheath thickness. The curves illustrate the threshold nature of the opening condition and the approximately linear increase of the required threshold field with radio frequency.}
    \label{fig:angle-vs-Bp}
\end{figure}

Figure~\ref{fig:angle-vs-Bp} shows the collisionless opening half-angle as a function of axial surface dipole field for several radio frequencies. At each frequency, the opening angle is absent below the threshold stated by Eq.~\eqref{eq:window_existence}, or equivalently by Eq.~\eqref{eq:Bmin}, then grows rapidly at first and more gradually as it approaches large values. The open side of the threshold in Fig.~\ref{fig:angle-vs-Bp} is the region above and to the right of each threshold marker: once $B_p$ exceeds $B_{p,\min}$, the corresponding curve gives a real positive opening angle. The threshold field is proportional to frequency, so higher-frequency links require proportionally stronger magnetic fields to open a collisionless dipole window. This trend is opposite to the usual unmagnetized cutoff advantage of higher frequency, where increasing $f$ raises the critical electron density. In the magnetic-window model, frequency therefore plays a dual role: it makes the unmagnetized plasma less likely to be overdense, but it also makes cyclotron-assisted transmission harder to obtain for a fixed magnetic field because the condition is based on $\Omega_e\cos\psi>\omega$.

The same result also demonstrates the importance of interpreting field strength at the correct location. The relevant condition is not simply that the cyclotron frequency exceed the radio frequency at the vehicle surface. For a dipole source, the field must remain large enough at the outer boundary of the sheath, where it has been reduced by the factor $(1+t_s/a)^{-3}$. Consequently, a surface field that appears adequate under a uniform-field estimate may fail once finite sheath thickness is included.

Near the opening threshold, the angular aperture is especially sensitive to small changes in field strength. The expansion leading to Eq.~\eqref{eq:near_threshold_angle} shows that a magnetic source only marginally above threshold produces a narrow cone, so a substantially larger margin in $Y_p$ is needed to obtain a broad polar cap. This observation is significant for practical design because magnetic-field generation on a compact, high-speed body is constrained by mass, power, heating, structural integration, and, for pulsed systems, timing and energy-storage limitations.

The collisionless scaling also shows that larger projectiles are favored when the sheath thickness is fixed. Increasing $a$ reduces $\Lambda=t_s/a$, which weakens the cubic dipole penalty. Conversely, a small projectile surrounded by a sheath whose thickness is comparable to its radius pays a severe field-decay penalty. This is one of the main differences between an onboard dipole source and an externally imposed uniform magnetic field. The onboard dipole geometry naturally creates a polar magnetic window, but the aperture is strongly constrained by the finite distance over which the field must remain effective.

\subsection{Validation Against Full Cold-Plasma Dispersion Roots}
\label{subsec:appleton-validation}

The reduced opening criterion used in the dipole model is the projected-cyclotron condition in Eq.~\eqref{eq:projected_cyclotron_condition}, or equivalently $Y\cos\psi>1$ with $Y=\Omega_e/\omega$. This condition is a simplified representation of the right-hand or whistler-like branch of the cold magnetized-plasma dispersion relation. To check the accuracy of this reduction, the present subsection compares it with numerical roots of the full cold-plasma dispersion relation for oblique propagation in a uniform magnetized plasma.

The comparison is local rather than global. It does not include the dipole geometry, sheath thickness, or collisional attenuation. Instead, it tests the reduced angular condition at a point in the sheath by asking whether the same angular boundary is obtained from the complete cold-plasma tensor. This is the appropriate validation step because the dipole opening-cone formula is obtained by applying the local condition along a radial escape path.

For a cold electron plasma with background magnetic field along the local $z$ direction, the dielectric tensor and the corresponding quadratic dispersion relation have already been given in Eqs.~\eqref{eq:cold_tensor}--\eqref{eq:cold_C}. In this subsection the roots are evaluated after rewriting the response in terms of the dimensionless parameters $X=\omega_p^2/\omega^2$ and $Y=\Omega_e/\omega$. The quantities referred to below as roots are the two allowed values of $n^2$, i.e. the squared refractive indices of the two local electromagnetic normal modes, not separate roots of a scalar permittivity. The two roots of Eq.~\eqref{eq:cold_quadratic} are
\begin{equation}
    n^2_{\pm}
    =
    \frac{
    B\pm\left(B^2-4AC\right)^{1/2}
    }{2A}.
\end{equation}

The right-hand branch is identified by continuity from the parallel-propagating root. At $\psi=0$, the two roots reduce to
\begin{equation}
    n^2=R
    \qquad
    \mathrm{and}
    \qquad
    n^2=L.
\end{equation}
For $Y>1$, the right-hand root satisfies
\begin{equation}
    R
    =
    1+\frac{X}{Y-1},
\end{equation}
and is positive even when $X>1$. This is the local cold-plasma expression of the magnetic-window branch. The numerical comparison therefore tracks the root that begins at $n^2=R$ when $\psi=0$ and determines the largest angle for which the tracked root remains real and positive.

\begin{figure}[t]
    \centering
    \includegraphics[width=0.74\linewidth]{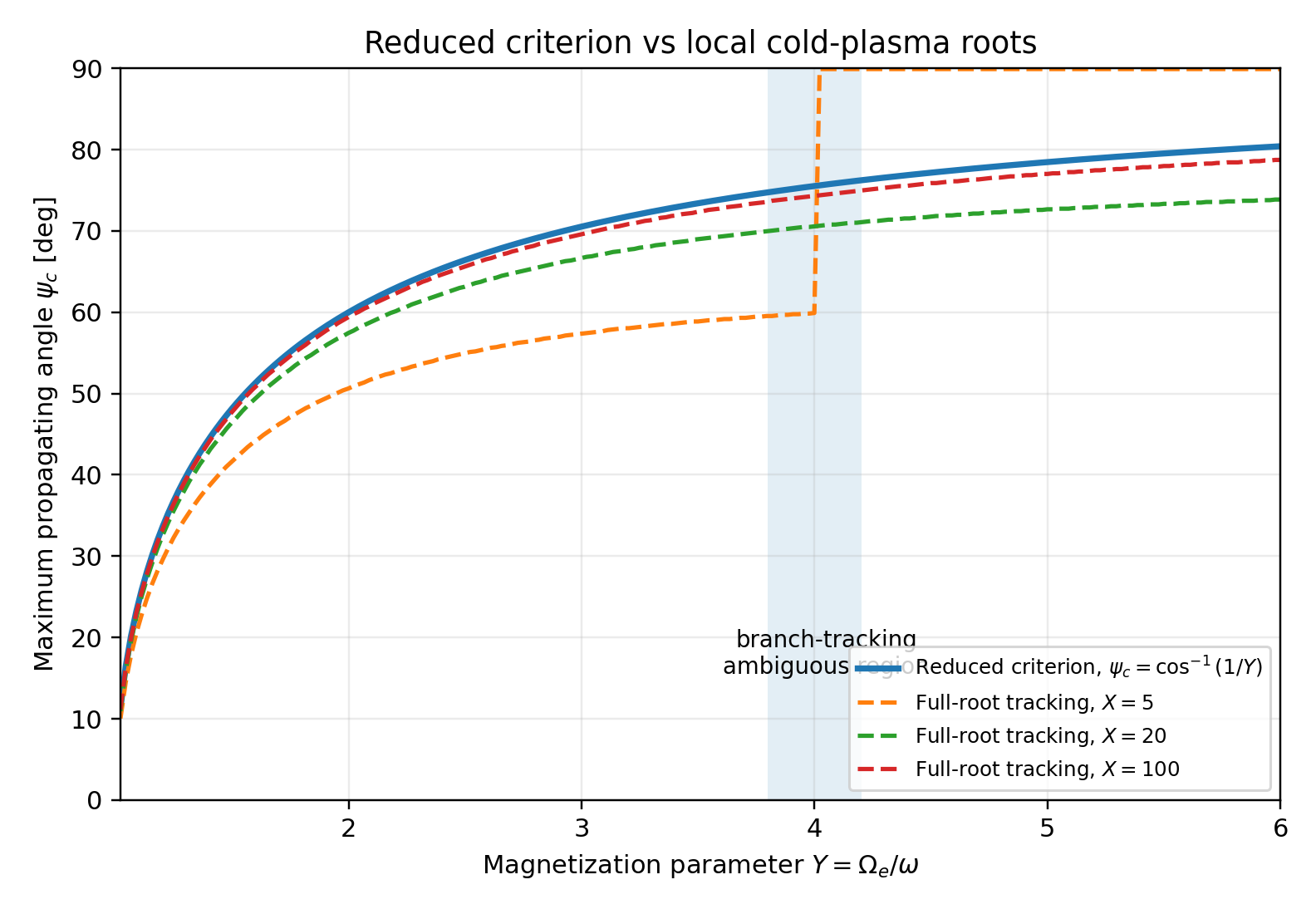}
    \caption{Comparison between the reduced magnetic-window angular condition $\psi_c=\cos^{-1}(1/Y)$ and the allowed $n^2$ roots of the full local cold-plasma dispersion relation for oblique propagation, where $Y=\Omega_e/\omega$ and $X=\omega_p^2/\omega^2$. The full-root calculation tracks the branch continuous with the parallel right-hand mode and determines the largest angle for which that tracked branch remains propagating. The shaded band marks a region near $Y\simeq4$ where the simple nearest-root tracking used for the $X=5$ diagnostic becomes ambiguous; the comparison is therefore a local consistency check, not a full validation of finite-thickness, collisional, inhomogeneous transmission.}
    \label{fig:appleton-validation}
\end{figure}

The reduced criterion predicts the angular boundary
\begin{equation}
    \psi_c^{\rm red}
    =
    \cos^{-1}\left(\frac{1}{Y}\right).
    \label{eq:validation_reduced_angle}
\end{equation}
The full-root calculation instead obtains $\psi_c$ by solving Eq.~\eqref{eq:cold_quadratic} numerically as $\psi$ is varied and following the right-hand branch until the propagating condition is lost. Figure~\ref{fig:appleton-validation} compares the reduced prediction with the full cold-plasma roots for several values of $X$.

The comparison suggests that the reduced projected-cyclotron criterion captures the main angular boundary of the cold-plasma right-hand branch over the overdense cases tested here. The visible deviation in the $X=5$ curve near $Y\simeq4$ occurs where the two local roots approach closely enough that the simple nearest-root continuation can switch branches. This feature should not be interpreted as a physical discontinuity in transmission through a plasma sheath. It is a diagnostic limitation of the local branch-tracking procedure. For the more overdense cases, $X=20$ and $X=100$, the tracked branch follows the reduced boundary more smoothly over the displayed range.

This calculation is therefore described only as a local consistency check. It is collisionless, uniform, and pointwise; it does not include spatial gradients, finite sheath thickness, reflection, tunneling, antenna coupling, or plasma-flow feedback. Its limited role is to support the use of $\Omega_e\cos\psi>\omega$ as the local propagation criterion from which the dipole opening-cone formula is derived. Collisional absorption and finite-thickness effects are treated separately in the following subsection.

\subsection{Loss-Limited Opening-Cone Scaling}

The collisionless opening angle derived in the preceding subsection gives the angular region in which the reduced magnetized-plasma branch can propagate in principle. It does not by itself determine whether the transmitted signal passes the assigned optical-depth screen. In a partially ionized sheath, electron-neutral and electron-ion collisions make the refractive index complex and reduce the amplitude of a wave traversing the sheath. The loss-limited model therefore replaces the condition of merely having a real opening angle with the stronger requirement that the optical depth remain below an assigned tolerance.

The reduced optical-depth model is Eq.~\eqref{eq:tau_reduced}, with detuning defined by Eq.~\eqref{eq:detuning}. For the radial escape approximation in an axial dipole field, the detuning at the outer edge of the sheath is given by Eq.~\eqref{eq:outer_detuning}. The loss-limited cone is then defined by requiring $\tau\leq\tau_*$, where $\tau_*$ is the maximum acceptable optical depth. Solving Eq.~\eqref{eq:tau_reduced} gives the minimum detuning in Eq.~\eqref{eq:delta_min}, and the corresponding loss-limited opening half-angle is Eq.~\eqref{eq:theta_loss_limited}.

Equation~\eqref{eq:theta_loss_limited} shows that collisional attenuation reduces the candidate loss-limited aperture by requiring additional magnetic detuning beyond the collisionless propagation condition. In the limit $\Delta_{\min}=0$, the loss-limited angle reduces to the collisionless angle. When $\Delta_{\min}>0$, the argument of the inverse cosine increases and the candidate loss-limited angle decreases. If the argument becomes greater than or equal to unity, the model predicts that no candidate loss-limited cone exists at the specified optical-depth tolerance.

This behavior highlights a distinction between a formally open magnetic window and a candidate loss-limited transmission path. A branch may be propagating according to the real part of the dispersion relation, yet still be too strongly attenuated to support a link. The severity of this effect is governed primarily by the product of electron density, collision frequency, and sheath thickness. Since $\omega_p^2$ is proportional to $n_e$, Eq.~\eqref{eq:tau_reduced} implies that dense plasma sheaths are intrinsically more lossy. Since $\nu_e$ appears explicitly in the numerator, weakly ionized but neutral-rich sheaths can also be strongly absorbing. The field strength enters through $\Delta$, so increasing $B_p$ can reduce loss by moving the wave farther from the boundary $\Delta=0$.

The loss-limited model also shows why operation near the opening threshold is undesirable. Close to threshold, the detuning is small, and the denominator $\Delta^2+\nu_e^2$ is not large enough to suppress collisional damping. A design that barely satisfies the collisionless condition may therefore produce a cone that is mathematically open but physically lossy. A more robust window requires field strength above the collisionless threshold by a margin determined by the plasma density, collision frequency, sheath thickness, and link budget.

The optical-depth tolerance $\tau_*$ should be interpreted as a placeholder for communication-system requirements. A smaller value of $\tau_*$ corresponds to a stricter amplitude-loss constraint and therefore a smaller candidate loss-limited cone. A larger value corresponds to a more permissive link budget. In an actual system, this tolerance would depend on transmitter power, antenna gain, receiver sensitivity, modulation, coding, background noise, and acceptable outage probability. The present model does not attempt to resolve these communication-system details; it uses $\tau_*$ only to translate collisional plasma loss into a geometric aperture.

\subsection{Dependence on Ambient Neutral Density and Projectile Speed}

Ambient neutral density and projectile speed enter the model through the reduced plasma-sheath closure. The upstream density $n_\infty$ controls the number of particles available for ionization and collision, while the speed $V$ controls the effective post-shock temperature used in the ionization estimate. The temperature, density, and collision-frequency expressions are given by Eqs.~\eqref{eq:Ts_model}, \eqref{eq:ne_chi}--\eqref{eq:nn_chi}, and \eqref{eq:nu_scaling}. These expressions make the density and speed dependencies qualitatively transparent, even though the closure is not intended as a validated hypersonic chemistry model.

\begin{figure}[t]
    \centering
    \includegraphics[width=0.74\linewidth]{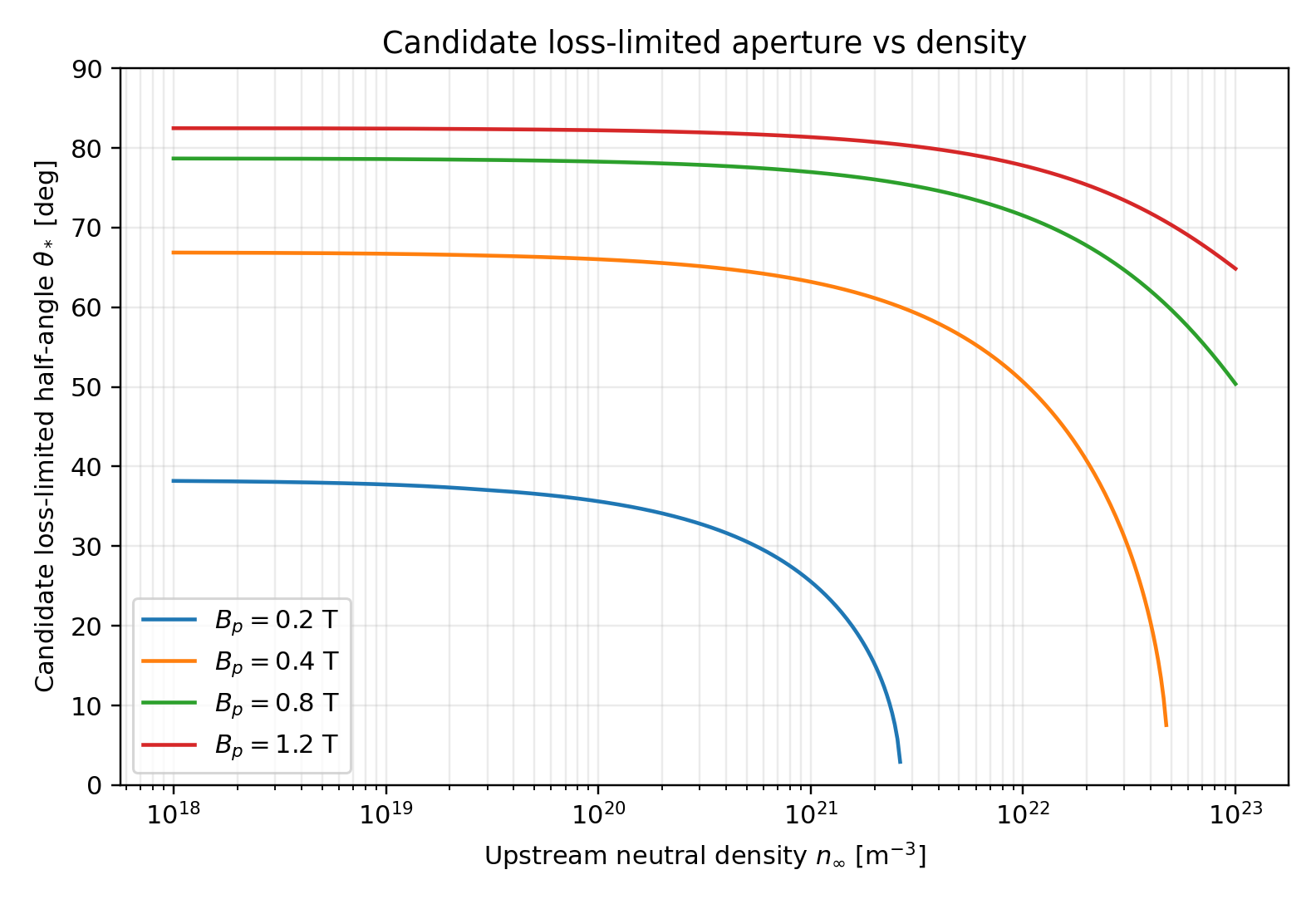}
    \caption{Candidate loss-limited magnetic-window half-angle as a function of upstream neutral density $n_\infty$ for several axial surface dipole fields, using the reduced ionization and collision-frequency closure. Increasing density generally raises the electron density and collision frequency, reducing the candidate loss-limited aperture even when the collisionless opening condition remains satisfied.}
    \label{fig:lossy-angle-vs-density}
\end{figure}

Figure~\ref{fig:lossy-angle-vs-density} shows the loss-limited opening angle as a function of ambient neutral density for several axial surface dipole fields. At low density, the sheath may be weakly ionized enough that the unmagnetized cutoff condition is not severe, and the collisional optical depth can also remain modest. As density increases, the model generally predicts greater electron density and greater collision frequency, both of which increase attenuation. The loss-limited cone therefore narrows and can disappear even when the collisionless magnetic-window condition is still satisfied.

The density dependence illustrates an important tradeoff. A denser atmosphere can make plasma blackout more severe by raising the electron density, but it also increases collisional absorption by raising the neutral density when the gas is not fully ionized. These two effects are not identical. The cutoff condition depends primarily on $n_e$ through $\omega_p$, whereas collisional damping depends on both $n_e$ and $\nu_e$. Consequently, a magnetic window may remain geometrically open while the candidate loss-limited region shrinks because the sheath has become more absorbing.

\begin{figure}[t]
    \centering
    \includegraphics[width=0.74\linewidth]{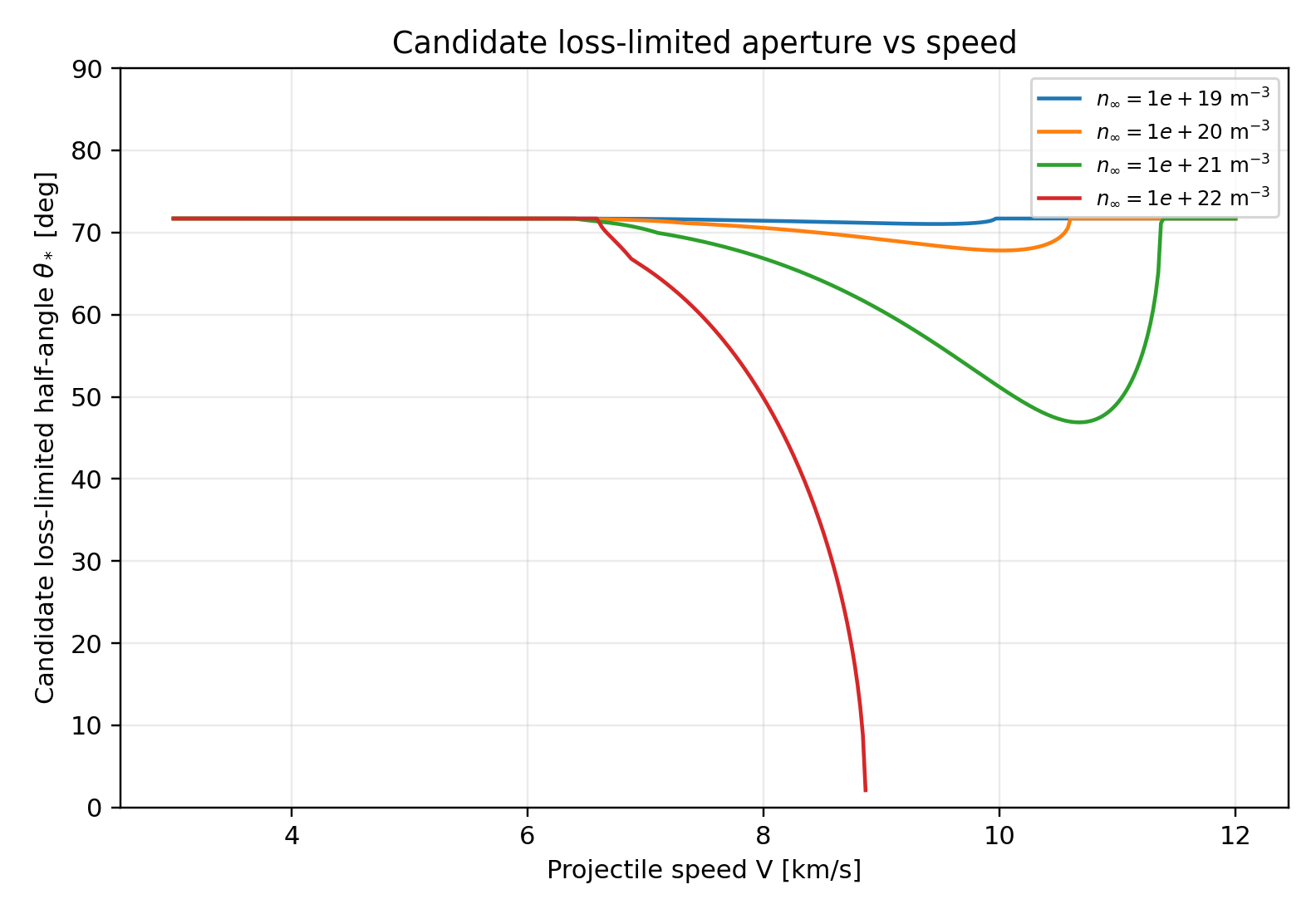}
    \caption{Candidate loss-limited magnetic-window half-angle as a function of projectile speed for several ambient neutral densities, using the reduced post-shock temperature and Saha-type ionization closure. Increasing speed raises the effective sheath temperature and can sharply increase the electron density, thereby increasing collisional optical depth and reducing the candidate loss-limited aperture for fixed magnetic field strength and radio frequency.}
    \label{fig:lossy-angle-vs-speed}
\end{figure}

Projectile speed enters in a different way. Increasing $V$ raises the effective post-shock temperature in Eq.~\eqref{eq:Ts_model}. Through the Saha factor, this can sharply increase the ionization fraction once the temperature reaches the range where the exponential ionization term becomes significant. Figure~\ref{fig:lossy-angle-vs-speed} shows the loss-limited opening angle as a function of projectile speed for several upstream neutral densities. At lower speeds, the model may predict weak ionization and modest plasma loss. As speed increases, the electron density can rise rapidly, causing the optical depth to increase and the candidate loss-limited cone to narrow. At sufficiently high speed, the ionization fraction may approach its cap in the reduced model, after which the trend depends on the remaining neutral density and the assumed collision mechanism.

\begin{figure}[t]
    \centering
    \includegraphics[width=0.74\linewidth]{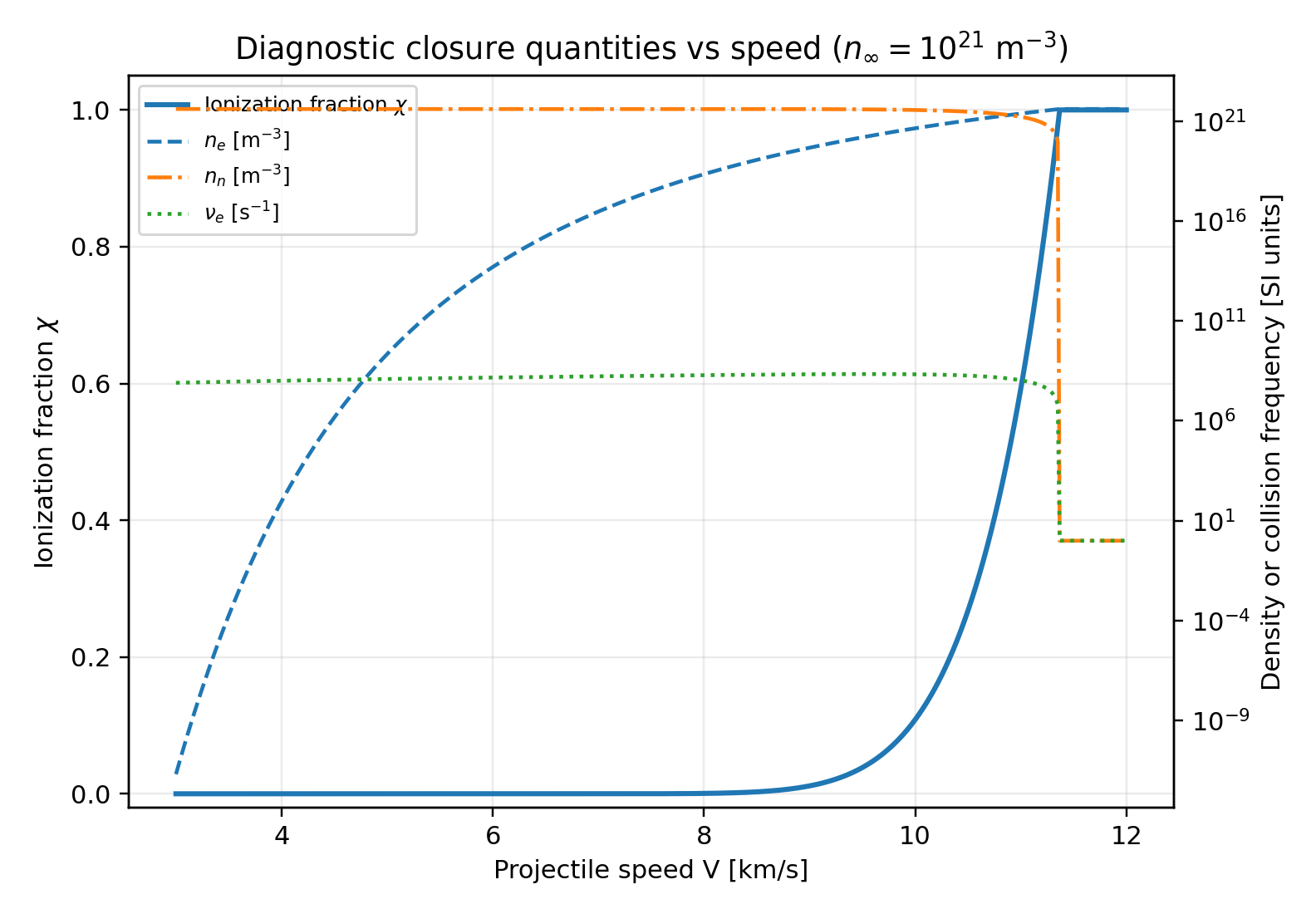}
    \caption{Diagnostic quantities from the reduced sheath closure as functions of projectile speed for $n_\infty=10^{21}~\mathrm{m^{-3}}$. The figure shows the ionization fraction $\chi$, electron density $n_e$, neutral density $n_n$, and electron-neutral collision frequency $\nu_e$ used to interpret the speed-dependent aperture curves. Very small low-speed electron densities are formal outputs of the illustrative equilibrium closure and are not intended to represent communication-relevant flight plasmas. The trends are closure diagnostics rather than validated nonequilibrium plasma-chemistry predictions.}
    \label{fig:speed-closure-diagnostics}
\end{figure}

Figure~\ref{fig:speed-closure-diagnostics} shows the reduced closure quantities that underlie the speed-dependent aperture curves for the representative case $n_\infty=10^{21}~\mathrm{m^{-3}}$. The ionization fraction rises rapidly once the effective temperature becomes large enough for the Saha factor to increase, while the remaining neutral density and electron-neutral collision frequency change as the gas approaches the imposed ionization cap. The non-monotonic portions of Fig.~\ref{fig:lossy-angle-vs-speed} therefore arise from the interaction of increasing electron density, decreasing neutral density, and the optical-depth dependence on $\omega_p^2\nu_e/(\Delta^2+\nu_e^2)$. Because these trends are produced by the simplified closure, they are interpreted as diagnostic behavior of the reduced model rather than as quantitative flight predictions. In particular, the very small electron densities at the low-speed end of the diagnostic plot simply indicate that the reduced equilibrium closure has not yet ionized the gas appreciably; they should not be interpreted as realistic communication-band plasma densities. In particular, the very small electron densities at the low-speed end of the diagnostic plot simply indicate that the reduced equilibrium closure has not yet ionized the gas appreciably; they should not be interpreted as realistic communication-band plasma densities.

The trends in Figs.~\ref{fig:lossy-angle-vs-speed} and~\ref{fig:speed-closure-diagnostics} should be interpreted as qualitative rather than predictive. The simplified temperature model uses a single parameter $\eta$ to represent real-gas effects, and the Saha closure assumes equilibrium behavior that may not hold in a rapidly evolving hypersonic sheath. Nevertheless, the result captures a physically plausible sequence. Firstly, increasing speed makes ionization more likely. Secondly, increased ionization raises the plasma frequency and therefore the strength of the plasma response. Thirdly, the resulting sheath can become more difficult to penetrate unless the magnetic detuning is increased by a stronger field or a more favorable geometry.

Taken together, the density and speed results show that the magnetic-window problem is not governed by magnetic field strength alone. A field that opens a broad collisionless cone in the reduced dilute-sheath model may fail to produce a candidate link in a denser or more strongly ionized sheath because collisional absorption increases the required detuning. Conversely, a high-frequency link may avoid unmagnetized cutoff at moderate electron density but require a larger magnetic field to satisfy the cyclotron-based opening condition. These competing trends motivate the regime-map analysis developed in the later results section.

\subsection{Dependence on Sheath Thickness, Projectile Scale, and Radio Frequency}

The finite thickness of the plasma sheath enters the dipole-field model through the dimensionless ratio $\Lambda=t_s/a$ defined in Eq.~\eqref{eq:Yp_Lambda}. This ratio is important because the magnetic source is assumed to be carried by the projectile. Unlike an externally imposed uniform field, a projectile-borne dipole field decreases rapidly with distance from the body. The field strength at the outer edge of the sheath is reduced by the factor $(1+t_s/a)^{-3}=(1+\Lambda)^{-3}$, and the corresponding collisionless opening angle is Eq.~\eqref{eq:theta_open_dimensionless}. This expression indicates that sheath thickness and projectile scale cannot be treated independently. A sheath of fixed physical thickness is less penalizing for a larger projectile, while the same sheath thickness can be prohibitive for a smaller projectile.

\begin{figure}[t]
    \centering
    \includegraphics[width=0.74\linewidth]{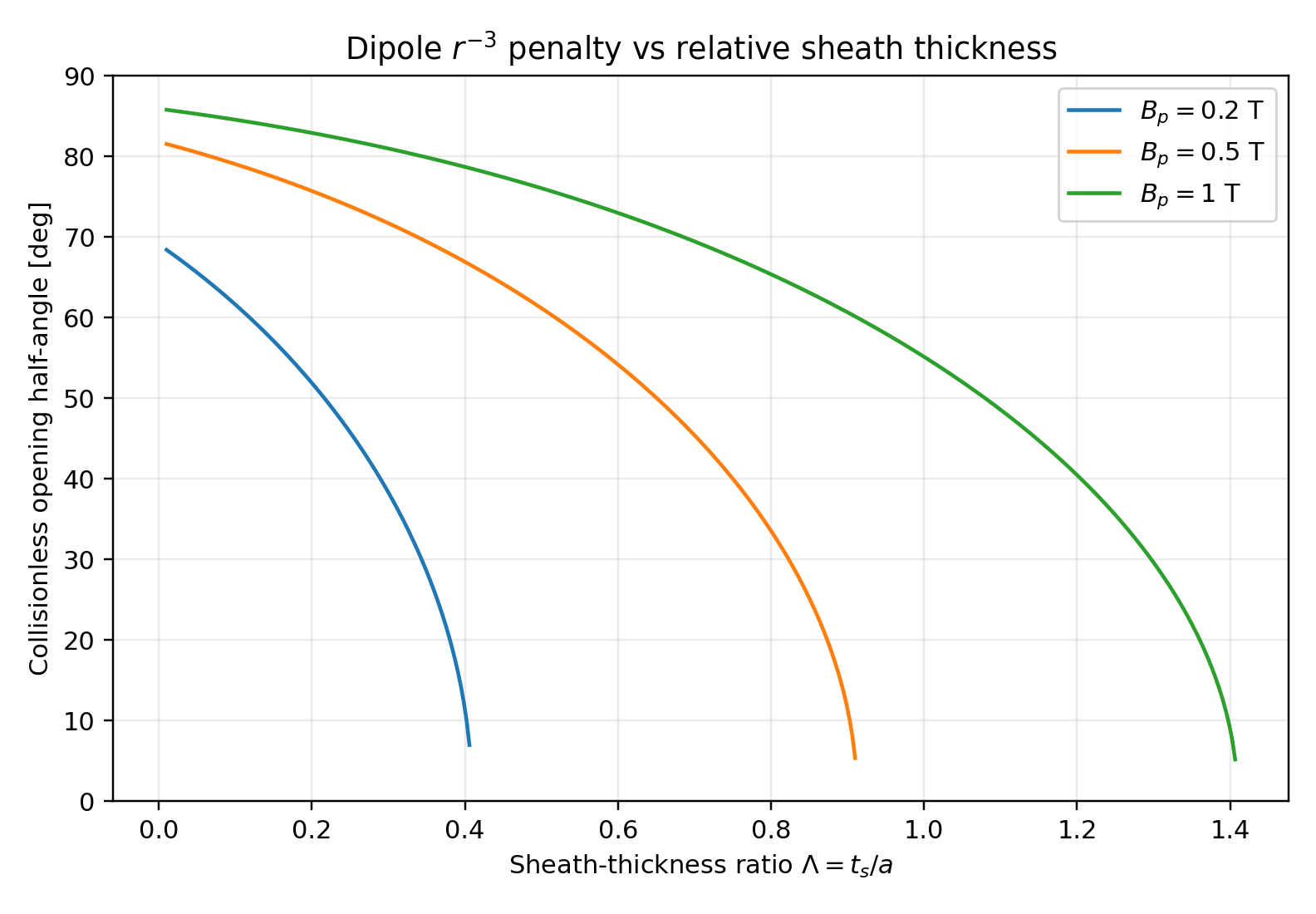}
    \caption{Collisionless magnetic-window opening half-angle as a function of the dimensionless sheath thickness $\Lambda=t_s/a$ for several axial surface dipole fields. The rapid decrease of the opening angle with $\Lambda$ reflects the $r^{-3}$ falloff of the projectile-borne dipole field, which requires the magnetic field to remain strong enough at the outer boundary of the sheath rather than only at the vehicle surface.}
    \label{fig:angle-vs-sheath-ratio}
\end{figure}

Figure~\ref{fig:angle-vs-sheath-ratio} is generated directly from Eq.~\eqref{eq:theta_open_dimensionless} by varying $\Lambda=t_s/a$ while holding the radio frequency fixed and using separate curves for the axial surface field. Moving to the right on the horizontal axis corresponds either to increasing the sheath thickness at fixed vehicle scale or to decreasing the vehicle scale at fixed sheath thickness. Each curve terminates when the inverse-cosine argument reaches unity, equivalently when $Y_p=(1+\Lambda)^3$. The termination is therefore the analytical opening threshold, not a numerical artifact. The rapid closing of the aperture follows from the $r^{-3}$ dipole falloff: the projected cyclotron frequency must remain above the radio frequency at the outer sheath boundary rather than only near the surface. This figure is less dependent on the Saha-type plasma closure than the density and speed plots because it follows mainly from the assumed dipole geometry and radial escape approximation.

The same scaling may be interpreted in terms of projectile size. At fixed sheath thickness $t_s$, increasing the characteristic radius $a$ decreases $\Lambda$ and therefore widens the opening cone. At fixed projectile scale, increasing the sheath thickness narrows the cone and may close it altogether. This result suggests that magnetic-window feasibility is likely to depend strongly on the ratio of sheath thickness to vehicle scale, not only on the absolute field strength. A field that is adequate for a large reentry body may be inadequate for a smaller projectile if the plasma layer occupies a larger fraction of the projectile radius.

Radio frequency enters through $Y_p=eB_p/(m_e2\pi f)$ in Eq.~\eqref{eq:Yp_Lambda}. At fixed magnetic field, increasing $f$ decreases $Y_p$ and therefore narrows the magnetic-window cone. The minimum field required to open any collisionless cone is given by Eq.~\eqref{eq:Bmin}; thus the magnetic-window requirement becomes more demanding at higher radio frequency. This trend contrasts with the unmagnetized cutoff condition in Eq.~\eqref{eq:necrit_scaling}, for which the critical electron density scales as $f^2$. Higher frequency makes an unmagnetized plasma less likely to be overdense, but it also requires a stronger magnetic field to satisfy the cyclotron-based opening condition. This dual role means that the optimal frequency for a magnetic-window communication system is not obvious from either criterion alone.

The reduced model therefore suggests a tradeoff. Lower frequencies are easier to support magnetically because the required cyclotron frequency is smaller, but they are more vulnerable to ordinary plasma cutoff because $n_{e,\mathrm{crit}}$ is smaller. Higher frequencies are more robust against unmagnetized cutoff, but they demand stronger magnetic fields for whistler-like transmission through a dipole window. In a full system design, this tradeoff would be resolved by combining plasma density profiles, magnetic-field limits, antenna coupling, receiver sensitivity, and acceptable link attenuation.

Sheath thickness also affects the loss-limited aperture beyond the collisionless dipole penalty, because the optical-depth estimate in Eq.~\eqref{eq:tau_reduced} contains an explicit factor of $t_s$. A thicker sheath therefore reduces transmission in two ways. Firstly, it places the outer boundary farther from the dipole source, reducing magnetic detuning through the factor $(1+\Lambda)^{-3}$. Secondly, it increases the path length over which collisional absorption occurs. These two effects reinforce each other, making sheath thickness one of the most important parameters in the reduced model.

The implications for scaling are direct. A favorable magnetic-window regime is more likely when the projectile is large relative to the sheath thickness, the surface magnetic field is large compared with $m_e\omega/e$, and the radio frequency is not so high that the cyclotron condition becomes prohibitive. A favorable communication regime also requires that the selected frequency not be so low that the unmagnetized plasma cutoff and collisional absorption become overwhelming. The feasible design space is therefore bounded simultaneously by geometry, plasma density, field strength, and frequency.

\subsection{Regime Classification Maps}

The preceding subsections considered individual dependencies, but the magnetic-window problem is inherently multidimensional. A useful way to summarize the reduced model is to classify each point in parameter space according to whether the plasma sheath is not overdense, overdense without a candidate loss-limited transmission region, or overdense with a candidate loss-limited transmission region. This classification does not represent a full communication-link calculation, but it provides a compact map of the regimes predicted by the analytical model.

The first classification boundary is the unmagnetized blackout condition. A point is classified as not locally overdense when $n_e\leq n_{e,\mathrm{crit}}$, with $n_{e,\mathrm{crit}}$ defined in Eq.~\eqref{eq:necrit_scaling}. In this regime, the elementary cold-plasma cutoff condition does not require a magnetic window. Propagation may still be affected by collisional damping, gradients, and antenna coupling, but the sheath is not opaque in the simplest unmagnetized cutoff sense.

The second regime occurs when $n_e>n_{e,\mathrm{crit}}$ but the loss-limited dipole condition fails. This may occur because the collisionless opening condition in Eq.~\eqref{eq:window_existence} is not satisfied, or because collisional damping raises the required detuning enough that the loss-limited opening angle is no longer real. In this regime, the reduced model predicts plasma blackout without a candidate loss-limited magnetic-window path.

The third regime occurs when the sheath is overdense in the unmagnetized sense, $n_e>n_{e,\mathrm{crit}}$, and the loss-limited aperture in Eq.~\eqref{eq:theta_loss_limited} is real. This is the regime of principal interest because the plasma would be opaque in the baseline unmagnetized model, yet the imposed dipole field creates a directional path that also satisfies the assigned optical-depth tolerance.

\begin{figure}[t]
    \centering
    \includegraphics[width=0.74\linewidth]{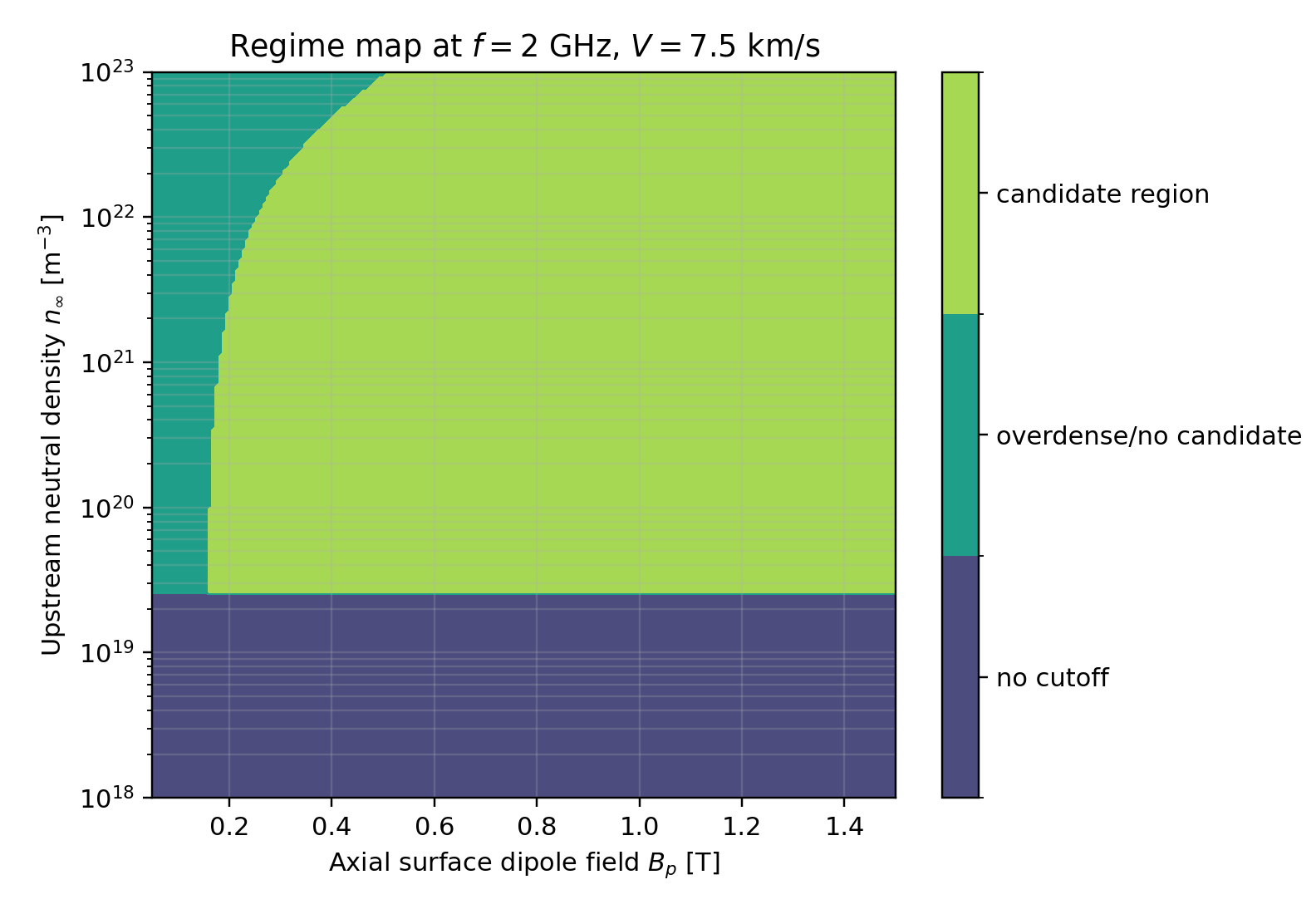}
    \caption{Example regime classification map in the plane of axial surface dipole field $B_p$ and upstream neutral density $n_\infty$ for fixed projectile scale, sheath thickness, radio frequency, and projectile speed. The reduced model distinguishes regions where the sheath is not overdense in the unmagnetized cutoff sense, where it is overdense without a candidate loss-limited transmission region, and where it is overdense but a candidate loss-limited dipole-field transmission region is predicted by the reduced model.}
    \label{fig:regime-map-Bp-density}
\end{figure}

Figure~\ref{fig:regime-map-Bp-density} shows an example regime map in the plane of axial surface dipole field and upstream neutral density. The map separates regions in which the reduced model predicts no unmagnetized blackout, blackout without a candidate loss-limited transmission region, and blackout with a candidate loss-limited transmission region. The boundaries should not be interpreted as sharp flight predictions, because the plasma closure and optical-depth model are simplified. They are best understood as organizing curves that show how the competing effects enter the feasibility problem.

The regime map illustrates, within the reduced model, the role of magnetic-field strength most clearly. At low field, increasing density can drive the sheath into the overdense blackout regime without providing any magnetic-window path. As the field increases, a region appears in which the magnetic detuning is large enough to support a candidate loss-limited cone. At still higher densities, collisional absorption and increasing plasma response can again reduce or eliminate the candidate loss-limited transmission region unless the field is increased further.

The map also illustrates why a single threshold field is not sufficient to characterize the problem. The collisionless threshold depends on frequency and sheath geometry, while the loss-limited threshold also depends on electron density, neutral density, collision frequency, and optical-depth tolerance. A field that opens a collisionless cone at one density may fail at another density because the attenuation requirement is more stringent. Conversely, a field that appears unnecessarily strong in a dilute sheath may be required in a denser sheath to maintain adequate detuning and suppress collisional loss.

Regime maps of this type are useful for selecting cases for higher-fidelity modeling. A full-wave or ray-tracing calculation is most valuable near the transition between blackout without a window and blackout with a candidate loss-limited transmission region, because the reduced model is most uncertain where gradients, mode conversion, and collision effects interact strongly. Similarly, nonequilibrium aerothermochemical simulation is most important near the transition between the no-blackout and blackout regimes, because the electron density predicted by the simplified closure may be especially sensitive to temperature, chemistry, and altitude.

The classification should therefore be viewed as a screening tool. It identifies parameter combinations where a projectile-borne dipole field is clearly inadequate, where it is likely unnecessary under the elementary cutoff criterion, and where it may plausibly enable transmission through an otherwise overdense plasma sheath. The latter region is the appropriate target for more detailed electromagnetic, plasma-chemistry, antenna, and experimental studies.

\subsection{Sensitivity to Plasma-Closure Parameters}
\label{subsec:closure-sensitivity}

The loss-limited results depend on closure parameters that are not determined by the reduced analytical model itself. The most important of these are the effective shock-heating factor $\eta$, the electron-neutral momentum-transfer cross section $\sigma_{en}$, and the relative sheath thickness $t_s/a$. The first parameter controls the effective post-shock temperature used in the Saha-type ionization estimate, the second controls the electron-neutral collision frequency, and the third controls both the dipole-field falloff and the absorption path length. A sensitivity calculation is therefore useful for separating robust geometric trends from quantities that depend on the simplified plasma closure.

The baseline case used for the sensitivity calculation is the same order-of-magnitude case used in the preceding parametric plots, with fixed projectile scale, radio frequency, surface dipole field, ambient density, and speed. Each uncertain parameter is then multiplied by a scalar factor while the other inputs are held at their baseline values. The result is not a statistical uncertainty analysis, since no probability distribution is assigned to the parameters. It is instead a local robustness test of the reduced-order aperture criterion.

\begin{figure}[t]
    \centering
    \includegraphics[width=0.74\linewidth]{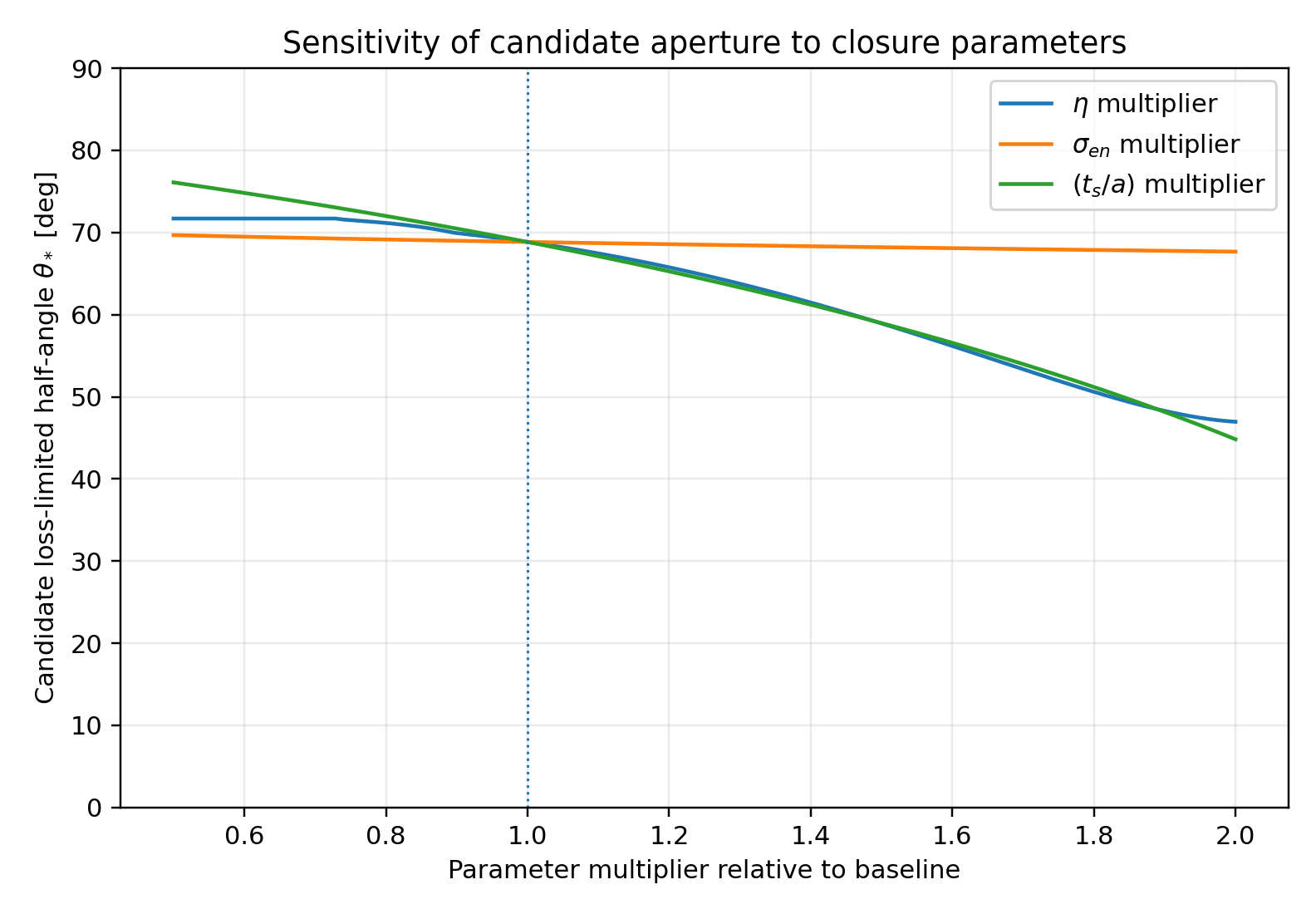}
    \caption{Sensitivity of the candidate loss-limited magnetic-window half-angle to uncertain closure parameters, expressed as multipliers relative to the baseline values used in the parametric model. The plotted parameters are the effective shock-heating factor $\eta$, the electron-neutral momentum-transfer cross section $\sigma_{en}$, and the relative sheath-thickness factor $t_s/a$. The result is intended as a robustness check for the reduced-order screening model rather than as a probabilistic uncertainty quantification.}
    \label{fig:sensitivity-closure-parameters}
\end{figure}

Figure~\ref{fig:sensitivity-closure-parameters} shows the resulting loss-limited opening half-angle as a function of the parameter multiplier. The strongest sensitivity in this example is associated with $t_s/a$, because the sheath thickness appears in two places. It enters the geometric dipole penalty through $(1+t_s/a)^3$, and it also enters the optical-depth estimate through the absorption path length. Increasing $t_s/a$ therefore narrows the candidate loss-limited cone by weakening the magnetic detuning at the sheath boundary and by increasing the distance over which collisional loss accumulates.

The sensitivity to $\sigma_{en}$ reflects the role of electron-neutral collisions in the optical-depth model. Increasing $\sigma_{en}$ raises the effective collision frequency, which generally increases the detuning required to keep the optical depth below the selected tolerance. The resulting effect on the candidate loss-limited cone is weaker than the sheath-thickness effect in the displayed baseline case, but it remains important because $\sigma_{en}$ may vary with species composition and electron energy. A more complete model should therefore replace the single effective cross section with species-dependent momentum-transfer data evaluated using the local electron-energy distribution.

The sensitivity to $\eta$ enters through the effective temperature in the ionization closure. Increasing $\eta$ raises the estimated post-shock temperature and can increase the ionization fraction through the exponential factor in the Saha-type expression. The effect is case dependent because the ionization fraction may remain small, rise rapidly, or approach the imposed cap depending on the selected density and speed. The reduced model should therefore not be used to make quantitative claims about speed thresholds without comparison to nonequilibrium plasma-chemistry calculations.

The sensitivity plot supports a modest interpretation of the numerical results. The collisionless aperture formula is a relatively robust geometric consequence of the dipole-field assumption and the projected cyclotron condition. The loss-limited aperture, however, is more model dependent because it requires estimates of electron density, neutral density, collision frequency, and sheath thickness. The results should therefore be interpreted as candidate regimes for follow-on simulation rather than as final transmission predictions. In particular, the parameter combinations near the boundary between a finite and absent candidate loss-limited cone are the most appropriate targets for full-wave propagation, ray tracing, or nonequilibrium aerothermochemical modeling.

\section{Comparison With Prior Magnetic-Window Studies}

The magnetic-window concept examined in this paper belongs to a broader class of plasma-blackout mitigation approaches in which the electromagnetic properties of the sheath are modified rather than merely bypassed by increasing transmitter power or changing antenna location. Prior work has shown that a magnetized plasma can support propagation regimes that differ qualitatively from the corresponding unmagnetized cutoff condition. In particular, right-hand and whistler-like modes can propagate preferentially along magnetic-field directions when the cyclotron response is sufficiently strong. The present study is consistent with that physical picture, but it does not seek to re-establish the existence of magnetic windows. Its narrower contribution is the derivation of a compact dipole-field scaling model for the angular aperture and candidate loss-limited aperture of such a window when the magnetic source is carried by the projectile.

Earlier analyses of reentry communication blackout emphasized the role of electron density, plasma frequency, and collisional attenuation in determining whether radio-frequency waves can penetrate the sheath \cite{Rybak1971, Kim2009Review}. These works motivate the baseline cutoff criterion associated with Eq.~\eqref{eq:unmag_dispersion}, as well as the need to treat absorption separately from cutoff. In this respect, the present model follows the standard interpretation of blackout as a combined reflection, cutoff, and absorption problem. The unmagnetized cutoff condition is used only as a reference state; it is not assumed to describe the full transmission problem in a realistic hypersonic plasma.

NASA-sponsored studies of electromagnetic propagation in magnetized reentry plasmas have already identified the possibility that imposed magnetic fields can create transmission windows through otherwise opaque plasma layers \cite{Manning2009}. Those studies examined the dependence of reflection, absorption, and transmission on magnetic-field strength, collision frequency, and plasma density, and they emphasized that the magnetic field may enable propagation through whistler-like modes. The present model is aligned with that mechanism. Its distinction is that it reduces the field geometry to an onboard axial dipole and asks how the finite sheath thickness and the $r^{-3}$ dipole decay constrain the angular region through which a wave may escape.

Numerical simulations of blackout and magnetic-window mitigation have also treated the coupled electromagnetic and plasma response using more detailed models than the reduced analysis developed here \cite{Kundrapu2015}. Such simulations are better suited to resolving spatial gradients, finite vehicle shape, plasma inhomogeneity, and time-dependent fields. The present work is not intended to replace those approaches. Instead, it provides a closed-form screening model that can be evaluated before high-fidelity simulation is attempted. The advantage of the analytical model is that it exposes parameter dependencies directly. The disadvantage is that it cannot capture mode conversion, gradient reflection, antenna coupling, or nonlinear plasma-flow feedback.

The closest overlap with the present study is found in work on wave propagation through hypersonic plasma sheaths magnetized by dipole magnetic fields \cite{Bai2022EHF}. That prior work demonstrates that dipole fields can materially affect electromagnetic-wave propagation in a sheath and is therefore directly relevant to the concept considered here. The present paper differs by focusing on a reduced aperture model rather than a detailed propagation calculation. The key analytical result in Eq.~\eqref{eq:theta_open_dimensional} makes explicit how a projectile-borne dipole field converts the magnetic-window condition into a polar opening angle controlled by surface field strength, radio frequency, projectile scale, and sheath thickness. This expression is not a substitute for the full propagation models in prior dipole-field studies, but it provides an interpretable scaling law that can be used to organize and preselect cases for such models.

Recent pulsed-field studies further show that magnetic manipulation of a plasma layer may create temporary or localized reductions in blackout severity \cite{Yuan2021Pulsed, Peng2025Pulsed}. These approaches are related in that they exploit electromagnetic control of the plasma environment, but their emphasis differs from the present steady or quasi-steady dipole-field aperture model. Pulsed-field methods may modify the density distribution or create transient transmission conditions, whereas the present analysis assumes a given sheath and asks whether a static or slowly varying dipole field can support a directional propagating branch with acceptable attenuation. A complete future treatment could combine these ideas by allowing $B_p$ and the plasma state to vary in time, thereby replacing the static opening angle with a time-dependent aperture.

The present study is therefore best viewed as an intermediate model between elementary cutoff estimates and full numerical simulations. Compared with elementary cutoff estimates, it includes anisotropic magnetized-plasma propagation, dipole geometry, and collisional attenuation. Compared with high-fidelity simulations, it intentionally omits detailed flow, chemistry, gradients, antenna coupling, and full-wave effects. Its novelty lies in the explicit scaling connection among projectile size, sheath thickness, dipole-field strength, radio frequency, and loss-limited angular aperture. This contribution is modest but useful: it gives a compact analytical framework for determining whether a projectile-borne magnetic-window concept is obviously infeasible, geometrically plausible but loss limited, or promising enough to justify more expensive modeling.

The comparison with prior work also clarifies the limitations of the claims made here. The model should not be interpreted as demonstrating that a full communication link can be maintained through a realistic hypersonic sheath. It demonstrates only that, under reduced cold-plasma and collisional assumptions, an onboard axial dipole field produces an analytically predictable polar aperture and that collisional damping can be incorporated as an optical-depth constraint. The next level of validation would require direct comparison with full-wave simulations, measured plasma-sheath profiles, and experimental data from controlled plasma facilities.

\section{Discussion}

The reduced model developed in this paper indicates that a projectile-borne axial dipole field can, in principle, create a directional magnetic window through an otherwise overdense plasma sheath. The result is not that the sheath becomes globally transparent. Rather, the imposed field changes the local dispersion relation so that a right-hand or whistler-like branch may propagate in a polar region where the wave vector is sufficiently aligned with the magnetic field. For the radial escape approximation, this condition reduces to a simple requirement on the radial component of the dipole field at the outer edge of the sheath.

The most important geometric result is the finite-sheath factor $(1+t_s/a)^3$ in Eq.~\eqref{eq:theta_open_dimensional}. This factor expresses the penalty associated with carrying the magnetic source on the projectile. The field must be strong enough not merely at the vehicle surface but at the outer boundary of the plasma sheath, where the wave exits into the surrounding medium. Since a dipole field decreases as $r^{-3}$, even a moderately thick sheath can substantially increase the required surface field. This effect is absent from uniform-field estimates and is therefore essential when assessing onboard magnetic-window concepts.

The collisionless opening angle gives an optimistic upper bound on the available aperture. It is controlled by the competition between the surface magnetization parameter $Y_p$ and the geometric penalty $(1+\Lambda)^3$, defined in Eq.~\eqref{eq:Yp_Lambda}. When $Y_p$ is less than this geometric penalty, no collisionless radial window exists in the reduced model. When $Y_p$ is only slightly larger, the opening angle is narrow. A broad aperture requires a field strength substantially above the minimum value in Eq.~\eqref{eq:Bmin}. This threshold relation is one of the most useful design estimates produced by the model.

Collisional damping imposes a second and often more restrictive condition. A wave may satisfy the real propagation criterion and still be attenuated strongly before leaving the sheath. The reduced optical-depth expression in Eq.~\eqref{eq:tau_reduced} indicates that the candidate loss-limited aperture is narrowed by electron density, collision frequency, and sheath thickness. It also shows why additional magnetic detuning can help. By increasing $\Delta$, a stronger magnetic field can reduce the factor $\nu_e/(\Delta^2+\nu_e^2)$ and thereby lower the absorption. Thus, magnetic-field strength has two roles: it opens the branch geometrically and moves the wave away from the most lossy boundary of the reduced dispersion relation.

The model also highlights a frequency tradeoff. Higher radio frequency increases the critical electron density for unmagnetized cutoff, which can make a plasma sheath less opaque in the baseline sense. At the same time, higher frequency increases the magnetic field required to satisfy the cyclotron-based opening condition. Lower frequency is easier to support magnetically, but more vulnerable to ordinary plasma cutoff and potentially stronger attenuation in an overdense sheath. A complete system analysis would therefore require frequency selection based on the combined plasma density, magnetic-field capability, antenna design, and link budget rather than on cutoff or cyclotron considerations alone.

Ambient density and projectile speed influence the result through the plasma state. In the simplified closure used here, increasing ambient density generally increases both the number of electrons available for plasma response and the number of neutrals available for electron-neutral collisions. Increasing projectile speed raises the effective post-shock temperature and can sharply increase ionization. These dependencies suggest that the magnetic-window aperture may vary strongly along a trajectory as altitude, speed, and sheath structure change. A field strength that is adequate in one portion of flight may be inadequate in another. This observation supports the possible value of adaptive frequency selection, pulsed magnetic fields, variable field strength, or trajectory-dependent communication scheduling in future studies.

The screening implications are mixed. The analytical model supports the physical plausibility of a dipole-field magnetic-window aperture under reduced assumptions, but it also exposes severe constraints. Small projectiles with relatively thick plasma sheaths are penalized by the dipole falloff. Dense or weakly ionized sheaths can be highly absorbing because collisions remain strong. Operation near the magnetic-window threshold may be fragile because the aperture is narrow and the detuning is small. These trends suggest that candidate feasibility is most favorable for cases in which the sheath is thin relative to the vehicle scale, the required radio frequency is not excessively high, the magnetic source can provide substantial surface field, and the plasma collision frequency is low enough that the loss-limited cone is not much smaller than the collisionless cone.

The present analysis also clarifies what must be added before the concept can be evaluated as an engineering design. Firstly, the plasma sheath should be computed using nonequilibrium aerothermochemistry rather than the reduced Saha-type closure used here. The electron density, neutral density, temperature, and collision frequency should be spatial profiles rather than single representative values. Secondly, electromagnetic propagation should be computed using the full complex magnetized-plasma dispersion relation or a full-wave solver, especially in regions where gradients, reflections, and mode conversion are important. Thirdly, antenna coupling must be included, because opening a whistler-like branch does not guarantee efficient excitation of that branch by a practical onboard antenna. Fourthly, the magnetic source itself must be modeled in terms of mass, volume, power, thermal load, structural integration, and possible interaction with the surrounding plasma flow.

The treatment of collisions should also be improved in future work. The present model uses an effective electron-neutral collision frequency and a perturbative estimate of the imaginary part of the refractive index. A more complete model should include electron-ion collisions, species-dependent momentum-transfer cross sections, nonequilibrium electron temperature, and the possibility of resonant absorption where the local cyclotron frequency approaches the radio frequency. In a spatially varying dipole field, the condition $\Omega_e\cos\psi\approx\omega$ may occur at particular locations in the sheath, potentially producing localized absorption layers. Such effects require integration through the full plasma and magnetic-field profiles.

The role of magnetic-field feedback on the flow is another important limitation. The present model treats the magnetic field as modifying only the radio-frequency wave response. If the field is sufficiently strong, however, it may also alter charged-particle transport, plasma density distribution, shock-layer structure, or sheath thickness. Such magnetohydrodynamic or kinetic feedback could either improve or degrade transmission depending on the configuration. The reduced model does not include these effects, so its predictions should be regarded as a first screening step rather than a final feasibility assessment.

Despite these limitations, the model may provide useful guidance. It identifies the combinations of field strength, frequency, vehicle scale, sheath thickness, and plasma state that are most worth studying with more expensive tools. It also provides simple expressions that can be used to check whether a proposed simulation or experiment lies in a plausible magnetic-window regime. If the reduced model predicts no collisionless opening angle, a detailed calculation may still reveal more complex effects, but the basic dipole-field concept is unlikely to be favorable. If the reduced model predicts a broad collisionless cone but a vanishing loss-limited cone, then collisional absorption is the central obstacle. If both the collisionless and loss-limited cones are finite, the case is a strong candidate for full-wave and experimental investigation.

The main conclusion from the discussion is therefore cautiously positive. A projectile-borne dipole field can create a mathematically well-defined magnetic-window aperture in a reduced magnetized-plasma model, and the aperture can be expressed by simple scaling laws. At the same time, the same scaling laws show that finite sheath thickness and collisional absorption can severely restrict the candidate aperture. The concept is therefore not established as an engineering communication solution by the present analysis, but it is sufficiently structured and physically plausible to justify targeted higher-fidelity modeling and laboratory validation.

\section{Model Limitations and Validation Requirements}

The model developed in this paper is intentionally reduced. Its purpose is to expose the leading scaling behavior of a projectile-borne dipole magnetic window, not to provide a validated prediction of communication performance for a specific vehicle, trajectory, or plasma environment. The analytical expressions derived above should therefore be interpreted as screening relations. They identify regimes that are geometrically plausible or implausible under the stated assumptions, but they do not replace full plasma-flow, electromagnetic, or communication-system analysis.

The first limitation is the use of a cold-plasma dispersion model. The cold-plasma tensor captures the essential anisotropy introduced by a magnetic field and gives the right-hand and whistler-like propagation behavior that motivates the magnetic-window concept. However, a hypersonic plasma sheath may contain strong gradients, multiple species, nonequilibrium electron temperatures, ionization and recombination reactions, vibrational excitation, dissociation, ablation products, and radiation. These effects can change both the real and imaginary parts of the refractive index. A higher-fidelity treatment should replace the local cold-plasma approximation with a complex, species-resolved plasma dielectric model evaluated using nonequilibrium sheath profiles.

The second limitation is the simplified plasma-sheath closure. The Saha-type ionization estimate and the effective shock-temperature model are useful for parametric exploration, but they are not adequate for quantitative flight prediction. In an actual hypersonic sheath, the electron density depends on the vehicle shape, altitude, speed, surface temperature, shock standoff distance, gas composition, finite-rate chemistry, and possible ablation or catalytic surface effects. The closure used here compresses these effects into a small number of parameters, especially $\eta$, $C_s$, and $E_i$. These parameters can be tuned to explore trends, but they cannot establish the electron-density field around a real projectile.

The third limitation is the treatment of the sheath as a layer with representative properties. The analytical model uses a characteristic sheath thickness $t_s$ and, in the loss estimate, characteristic values of $n_e$, $\nu_e$, and $n_r$. A real sheath has spatially varying electron density, neutral density, temperature, collision frequency, and magnetic-field strength. The appropriate attenuation measure is therefore an integral along a ray path,
\begin{equation}
    \tau
    =
    \frac{\omega}{c}
    \int n_i(s)\,ds,
\end{equation}
with $n_i(s)$ obtained from the local complex magnetized-plasma dispersion relation. The reduced model approximates this integral by a single effective value. That approximation is useful for scaling, but it cannot resolve localized absorption layers, evanescent tunneling regions, or reflection from sharp gradients.

The fourth limitation is the radial-ray assumption. The derivation of the dipole opening angle assumes that radio-frequency energy escapes approximately radially from the projectile. This assumption makes the projection factor analytically simple and leads to the result that the effective condition depends on the radial component of the dipole field. In a realistic anisotropic plasma, the ray direction and the group-velocity direction may differ, and energy may refract along paths that are not radial. Full treatment of this effect requires ray tracing through the Appleton--Hartree dispersion relation or direct full-wave simulation in the inhomogeneous magnetized sheath.

The fifth limitation is the treatment of the magnetic field as a vacuum dipole. The analysis assumes that the projectile-borne magnetic source produces an axial dipole field unaffected by the surrounding plasma. If the imposed field is strong enough, however, it may alter charged-particle transport, modify the sheath density distribution, influence shock-layer structure, or drive currents that distort the field. These effects may be beneficial or detrimental. They are not included in the present model, which treats the magnetic field as modifying the radio-frequency wave response but not the plasma-flow solution.

The sixth limitation is the perturbative treatment of collisional damping. The loss model assumes that the imaginary part of the refractive index is small compared with the real part and estimates the attenuation using an effective collision frequency. This approximation may fail near cyclotron-resonant regions, near mode cutoffs, or in strongly collisional plasma. A more complete treatment should solve for the complex refractive index directly using the full collisional dielectric tensor. It should also include electron-ion collisions, species-dependent electron-neutral cross sections, nonequilibrium electron temperature, and possible spatially localized resonant absorption where $\Omega_e\cos\psi$ approaches $\omega$.

The seventh limitation is the absence of an antenna model. The existence of a propagating magnetized-plasma branch does not guarantee that a physical antenna can efficiently excite it. The coupling depends on antenna location, orientation, polarization, impedance, near-field interaction with the plasma, and the spatial structure of the allowed mode. Since the magnetic-window branch is polarization- and direction-dependent, antenna design is central to practical feasibility. Future work should couple the sheath model to an antenna and full-wave electromagnetic solver to determine actual transmission coefficients rather than only geometric opening angles.

The eighth limitation is that no communication link budget is included. The optical-depth threshold $\tau_*$ is used as a proxy for acceptable loss, but a real communication system depends on transmitter power, receiver noise temperature, antenna gain, modulation format, coding, bandwidth, pointing, scintillation, and allowable outage time. A small aperture may still be useful if the link budget is favorable, while a wider aperture may be inadequate if coupling or noise conditions are poor. The present model therefore gives only a plasma-physics feasibility indicator, not a complete communication-system assessment.

Validation should proceed in stages. Firstly, the analytical opening-angle formula should be checked against numerical evaluation of the full cold-plasma dispersion relation for the same dipole geometry. This would test the accuracy of the reduced condition $\Omega_e\cos\psi>\omega$ over the range of angles and plasma densities of interest. Secondly, the loss-limited formula should be compared with direct integration of the complex refractive index through prescribed sheath profiles. This would identify where the representative-property approximation is adequate and where gradients dominate the result.

Thirdly, the plasma profiles used in the model should be replaced by nonequilibrium computational fluid dynamics or direct simulation data for representative hypersonic bodies. Such simulations should provide $n_e(\bm{x})$, $n_n(\bm{x})$, $T_e(\bm{x})$, $\nu_e(\bm{x})$, and sheath thickness as functions of altitude, speed, and vehicle geometry. The analytical model can then be evaluated using path-integrated quantities rather than a single effective layer. Fourthly, full-wave electromagnetic simulations should be performed in the resulting magnetized, inhomogeneous plasma to compute reflection, absorption, mode conversion, and transmission.

Experimental validation would require a controlled plasma environment in which density, collision frequency, magnetic field, and radio-frequency transmission can be measured. Arcjet facilities, plasma wind tunnels, shock tubes, and laboratory plasma devices could provide partial validation, although none perfectly reproduces the full hypersonic flight environment. A useful laboratory experiment would impose a known dipole or dipole-like magnetic field on a plasma layer, launch a polarized radio-frequency wave through the layer, and measure transmission as a function of field strength, frequency, density, and propagation angle. Such an experiment would directly test the predicted threshold and aperture trends.

The most important validation target is not the exact value of a single opening angle, but the scaling behavior. The model predicts a cubic sheath-thickness penalty, a linear frequency dependence of the minimum field, a widening aperture with increasing surface field, and a reduction of candidate loss-limited aperture due to collisional optical depth. If these trends are observed in full-wave simulation or experiment, the reduced model would be useful as a design and screening tool even if empirical correction factors are needed for quantitative prediction.

\section{Conclusions}

This paper has developed a reduced analytical screening model for radio-frequency transmission through a hypersonic plasma sheath in the presence of a projectile-borne axial dipole magnetic field. The model was motivated by the magnetic-window concept, in which magnetized-plasma anisotropy allows a right-hand or whistler-like mode to propagate along preferred directions even when the corresponding unmagnetized plasma would be overdense. The aim was not to introduce a new plasma-blackout mitigation mechanism, but to derive compact scaling relations for the angular aperture and candidate loss-limited aperture of a dipole-field magnetic window.

The principal collisionless screening relation is the polar opening half-angle in Eq.~\eqref{eq:theta_open_dimensional}, where $B_p$ is the axial surface dipole field, $f$ is the radio frequency, $a$ is the projectile scale, and $t_s$ is the sheath thickness. This expression indicates that the onboard dipole geometry imposes a strong finite-sheath penalty. Because the dipole field decays as $r^{-3}$, the field must remain strong enough at the outer sheath boundary, not merely at the vehicle surface.

The corresponding existence condition is Eq.~\eqref{eq:Bmin}, or equivalently the dimensionless criterion in Eq.~\eqref{eq:window_existence}. This condition summarizes, within the reduced model, the basic competition between magnetic-field strength, radio frequency, and sheath geometry. Stronger fields and larger projectile scales favor transmission, while higher frequencies and thicker relative sheaths close the window.

The model was then provisionally extended to include collisional absorption through the reduced optical-depth criterion in Eq.~\eqref{eq:tau_reduced}, with detuning defined in Eq.~\eqref{eq:detuning}. This expression indicates that a collisionless magnetic window is not necessarily a candidate transmission path. Dense, collisional, or thick sheaths can attenuate the wave strongly even when the propagating branch exists. A stronger magnetic field can help by increasing detuning, but the required field may exceed the collisionless threshold by a substantial margin.

The loss-limited aperture was expressed by Eq.~\eqref{eq:theta_loss_limited}, where $\Delta_{\min}$ is determined by the acceptable optical depth. This formula makes explicit, within the adopted approximation, the difference between a geometrically open cone and a candidate loss-limited cone. The former is controlled by cyclotron detuning and dipole geometry, while the latter is also controlled by electron density, collision frequency, and link-loss tolerance.

A simplified plasma-sheath closure was introduced to explore trends with ambient density and projectile speed. The closure used an effective post-shock temperature, a Saha-type ionization estimate, and an electron-neutral collision-frequency model. These assumptions are deliberately approximate, but they show how the magnetic-window aperture can shrink as electron density and collisional damping increase. The resulting parametric plots demonstrate the use of the model as a screening tool for identifying regions of no blackout, blackout without a candidate loss-limited transmission region, and blackout with a potentially candidate loss-limited transmission region.

The analysis supports a cautious conclusion. A projectile-borne axial dipole field can create a mathematically well-defined magnetic-window aperture in a reduced magnetized-plasma model, and the aperture obeys simple scaling laws. However, the same scaling laws reveal demanding constraints. The dipole falloff penalizes thick sheaths and small projectiles, higher radio frequencies require stronger magnetic fields for cyclotron-assisted propagation, and collisional absorption can substantially reduce the candidate loss-limited aperture. The concept is therefore physically plausible but not established as an engineering solution by the present analysis.

The most valuable next step is validation against higher-fidelity models. The reduced formulas should be compared with full Appleton--Hartree ray tracing, full-wave electromagnetic simulations, and nonequilibrium hypersonic plasma-sheath profiles. Laboratory experiments in controlled plasma devices with imposed dipole or dipole-like magnetic fields would also be useful for testing the predicted threshold and aperture scaling. If these studies confirm the main trends, the model could serve as a compact design tool for selecting magnetic-field strength, operating frequency, antenna placement, and candidate flight regimes for further investigation.

In summary, the paper proposes an analytical bridge between elementary plasma-cutoff estimates and detailed magnetic-window simulations. Its contribution is a transparent scaling framework for projectile-borne dipole fields, finite sheath thickness, and collisional loss. The framework identifies when a magnetic window is impossible under the reduced assumptions, when it is geometrically possible but absorption limited, and when it may be a candidate for full electromagnetic and aerothermochemical analysis.

\section*{Supplementary material}

Detailed algebraic derivations of the dipole opening-cone relation, the collisional optical-depth approximation, and the numerical implementation are provided as Supplementary Material. The main manuscript retains the essential equations and physical interpretation to reduce repetition while preserving the derivational path.

\section*{Data availability}
No external experimental datasets were used in this reduced-order study. The numerical data underlying the plotted curves are generated from the equations, parameter values, and modeling assumptions stated in the manuscript. The plotting code and generated numerical values can be made available upon reasonable request or supplied as supplementary material during submission.

\section*{Declaration of generative AI and AI-assisted technologies in the manuscript preparation process}
During preparation of this work, AI-assisted tools were used to support language editing, manuscript organization, LaTeX drafting, and preparation of an illustrative schematic figure. The author reviewed, edited, and verified the content as needed and takes full responsibility for the content of the manuscript.

\end{document}